\documentclass[twocolumn,secnumarabic,amssymb, nobibnotes]{revtex4}
\usepackage[T1]{fontenc}
\usepackage{color}
\usepackage{dcolumn}
\usepackage{textcomp}
\usepackage{graphicx}
\usepackage{revsymb}
\usepackage{subfigure}
\usepackage{amsbsy}
\usepackage{amsmath}
\usepackage{amsfonts}
\usepackage{amsthm}
\usepackage{float}
\usepackage{natbib}
\usepackage{epstopdf}

\theoremstyle{plain}

\begin{document}

\title{A study of Quantum Correlation for Three Qubit States under the effect of Quantum Noisy Channels}

\author{Pratik K. Sarangi$^{1,3}$, \& Indranil Chakrabarty$^2$} 
\affiliation{$^1$Center for Quantum Information \& Quantum Computation, Dept of Physics, Indian Institute of Science, Bangalore, India.\\
$^2$ Center for Security, Theory \& Algorithmic Research, International Institute for Information Technology, Hyderabad, India.\\ 
$^2$ The University of Arizona, Tucson, AZ 85721, United States }

\begin{abstract}
We study the dynamics of quantum dissension for three qubit states in various dissipative channels such as amplitude damping, dephasing and depolarizing. Our study is solely based on Markovian environments where quantum channels are without memory and each qubit 
is coupled to its own environment. We  start with mixed GHZ, mixed W, mixture of separable states, a 
mixed biseparable state, as the initial states and mostly observe that the decay of quantum dissension is asymptotic 
in contrast to sudden death of quantum entanglement in similar environments. This is a clear indication of the fact 
that quantum correlation in general is  more robust against the effect of noise. 
However, for a given class of initial mixed states  we find a temporary leap in quantum dissension for a certain 
interval of time. More precisely, we observe the revival of quantum correlation to happen for certain time period. 
This signifies that the measure of quantum correlation such as quantum discord, quantum dissension, defined from the information theoretic 
perspective is different from the correlation defined from the entanglement-separability paradigm and can increase 
under the effect of the local noise.  We also study the effects of these channels on the monogamy score of each of these initial states. 
Interestingly, we find that for certain class of states and channels, there is change from negative values to 
positive values of the monogamy score with classical randomness as well as with time. This gives us an important 
insight in obtaining states which are freely sharable (polygamous state) from the states which are not freely 
sharable (monogamous).  This is indeed a remarkable feature, as we can create monogamous states from polygamous states 
Monogamous states are considered to have more signatures of quantum ness and can be used for security purpose.    
\end{abstract}

\maketitle

\section{Introduction}
\noindent For a long time quantum entanglement was only of philosophical interest and researchers were mainly focusing on addressing the 
questions that were related with the quantum mechanical understanding of various fundamental notions like reality and locality 
\cite{Einstein1935}. However, for the last two decades world had seen that quantum entanglement is not 
only a philosophical riddle but also a reality as far as the laboratory preparation of entangled qubits are concerned 
\cite{Aspect1981}. Researches that were conducted 
during these decades were not all concerned about its existence but mostly about its usefulness as a resource to 
carry out information processing protocols like quantum teleportation \cite{Bennett1993}, cryptography \cite{Gisin2002}, superdense coding \cite{Bennett1992}, 
and in many other tasks \cite{Pati2001} . It was subsequently evident from various followed up investigations 
that quantum entanglement plays a pivotal role in all these information processing protocols. Therefore, 
understanding the precise nature of entanglement in bipartite and multiparty quantum systems has become the 
holy-grail of quantum information processing.

\noindent However, the precise role of entanglement as a resource in quantum information processing is  not fully understood 
and it was suggested that entanglement is not the only type of correlation present in quantum states. This is because lately some computational 
tasks were carried out even in the absence of entanglement \cite{KNILL}. This provided the foundation 
to the belief that there may be correlation present in the system even in the absence of entanglement. Hence, 
researchers redefined quantum correlation from the information theoretic perspective. This gave rise to various 
measures \cite{OLI, REST, IND, REST1} of quantum correlation, the predominant of 
them being quantum discord \cite{OLI}. Though there are issues that need to be addressed, in much deeper level 
quantum discord temporarily satisfies certain relevant questions. Subsequently, quantum discord has been given an operational interpretation 
in different contexts like quantum state merging \cite{Madhok2011} and remote state preparation \cite{Vedral2012}. 
In addition, extension of the notion of quantum discord to multi qubit cases has been proposed \cite{IND, REST1}.



\noindent Many works were done in the recent past to investigate the dynamics of quantum correlation in open systems by comparing 
the evolution of different types of initial states in specific models. These states are typically two qubits coupled 
with two local baths or one common bath. In principle, there are several factors that can affect the evolution, 
namely, the initial state for the system and environment, the type of system-environment interaction and the 
structure of the reservoir. A more relevant question will be how robust are these measures when they are subjected 
to the noise in quantum channels. 

\noindent It is mainly inspired by the studies of sudden death of entanglement for two qubits, having 
no direct interaction \cite{Eberly2004,Eberly2007}. Entanglement Sudden Death (ESD) is said to occur when 
the initial entanglement falls and remains at zero after a finite period of evolution for some choices of the initial 
state. ESD is a potential threat to  quantum algorithms and quantum information protocols and thus the quantum systems 
should be well protected against noisy environments. Another possible way to circumvent such resource vanishing is to make use of resources which do not suffer from sudden death.
At this point, one can ask a similar question: \textit{Does quantum discord present similar behavior?} In the 
first study \cite{Villas2009} addressing this question, researchers have compared the evolution of concurrence and 
discord for two qubits, each subject to independent Markovian decoherence (dephasing, depolarizing and amplitude 
damping). Looking at initial states such as Werner states and partially-entangled pure states, the authors find no 
sudden death of discord even when ESD does occur; quantum discord decays exponentially and vanishes asymptotically 
in all cases. However, not much is known about the effects on multipartite correlation with time when they are 
transferred through noisy quantum channels.

\noindent In this work, we study the dynamics of quantum dissension of three qubit states which happens to be a 
measure of multi party quantum correlation, under the effect of various 
quantum noisy channels. In addition, we also study the dynamics of monogamy score of these three qubit states in 
presence of channel noise. In section 2, we provide a detail descriptions of quantum dissension and monogamy score 
of quantum correlation. In section 3, we study the effect of various noisy channels on quantum correlation and 
monogamy score when all the qubits are transferred through them. Finally, we conclude in Section 4 by discussing future 
directions of explorations.

\section{Quantum Dissension \& Monogamy Score}
\noindent In classical information theory \cite{ThomasCover1991}, the total correlation between two random variables 
is defined by their mutual information. If X and Y are two random variables, the mutual information is obtained by 
subtracting the joint entropy of the system from the sum of the individual entropies. 
Mathematically, this can be stated as:
\begin{equation}
I(X:Y)= H(X)+H(Y)-H(X:Y),
\end{equation}
where $H(.)$ defines Shannon entropy function.
 
\noindent Another equivalent way of expressing mutual information is by taking into account the reduction in 
uncertainty associated with one random variable due to the introduction of another random variable. Stated formally as,
\begin{equation}
J(X:Y)=H(X)-H(X|Y),
\end{equation}
or
\begin{equation}
K(X:Y)=H(Y)-H(Y|X),
\end{equation}
where $H(X|Y)$ defines conditional entropy of $X$ given that $Y$ has already occurred and vice versa.

\noindent All these above expressions are equivalent in classical information theory. When we try to  quantify 
correlation in quantum systems from an information theoretic perspective, natural extension of these quantities 
will be obtained by replacing random variables with density matrices, Shannon entropy with Von Neumann entropy 
and apposite definition of the conditional entropies. Stated mathematically, the quantum mutual 
information is given by,
\begin{equation}
I(X:Y)=S(\rho_X)+S(\rho_Y)-S(\rho_{XY}),
\end{equation}
where $\rho_{XY}$ is the composite density matrix, 
$\rho_{X}$ and $\rho_{Y}$ are the local density matrices and $S(.)$ defines Von Neumann entropy function.

\noindent Similarly, by applying the argument of reduction of uncertainty associated with one quantum system with 
introduction of another quantum system, one can have the alternative definition of mutual information as,
\begin{equation}
J(X:Y)=S(\rho_{X})-S(\rho_{X|Y})
\end{equation}
and
\begin{equation}
K(X:Y)=S(\rho_{Y})-S(\rho_{Y|X}).
\end{equation}
\noindent Here $S(\rho_{X|Y})$ is the average of conditional entropy and is obtained after 
carrying out a projective measurement on subsystem $Y$ and vice versa. The projective measurement 
is done in the general basis
$\{|u_1\rangle$ = cos$\theta|0\rangle$ + $e^{i\phi}$ sin$\theta|1\rangle$,
  $|u_2\rangle$ = sin$\theta|0\rangle$ - $e^{i\phi}$ cos$\theta|1\rangle\}$,
where $\theta$ and $\phi$ have the range [0,2$\pi$].
Hence, the quantum conditional entropy can be expressed as,
$S(\rho_{X|Y})$ = $\sum_jp_jS(\rho_{X|\Pi_{jY}})$ where $p_j=tr[(I_X\otimes\Pi_{jY})\rho(I_X\otimes\Pi_{jY})]$,  
($I^{X}$ being identity operator on the Hilbert space of the quantum system $X$), gives the probability of 
obtaining the $j$the outcome. The corresponding post-measurement state of system 
$X$ is $\rho_{X|\Pi_{jY}}=\frac{1}{p_j} tr_{Y}[(I_X\otimes\Pi_{jY})\rho(I_X\otimes\Pi_{jY})]$. It is important to note over here that $S(\rho^{X|Y})$ is different from what will be the straightforward extension of 
classical conditional entropy. In quantum information, the meaning of conditional entropy of the qubit $X$ given that $Y$ has occurred 
is  the amount of uncertainty in the qubit $X$ given that a measurement is carried out on the qubit $Y$. 


\noindent Consequently, the expressions $I$, $J$ and $K$ are not equivalent in the quantum domain. 
The differences between $I-J$ and $I-K$ are captured by 
quantum discord i.e.
\begin{eqnarray}
D(X:Y)=I(X:Y)-J(X:Y) \nonumber\\
=S(\rho_Y)+\min_{\{\Pi_{jY}\}}S(\rho_{X|Y})-S(\rho_{XY}),
\end{eqnarray}
\begin{eqnarray}
D(Y:X)=I(X:Y)-K(X:Y) \nonumber\\
=S(\rho_X)+\min_{\{\Pi_{jX}\}}S(\rho_{Y|X})-S(\rho_{XY}).
\end{eqnarray}
\noindent  One variant 
of quantum discord is the geometric quantum discord which is defined as the distance between a quantum state and the nearest 
classical (or separable) state, \cite{Vedral2010}. Quantum discord has been established as a non-negative measure of correlation 
for any quantum states. Subsequent researches were carried out to obtain an analytical closed form of quantum discord and was found for certain class of states \cite{ARP2010,Luo2008}. An unified geometric view of quantum correlations which includes discord, entanglement along with the introduction of the concepts like quantum dissonance was given in \cite{modi}. 

\noindent One of the natural extension of quantum discord from two qubit to three qubit systems is quantum dissension \cite{IND}. 
Introduction of three qubits naturally brings in one and two-particle projective measurement into consideration. 
These measurements can be performed on different subsystems leading to multiple definitions of quantum dissension. 
In other words a single quantity is not sufficient enough to capture all aspects of correlation in multiparty systems.
Quantum dissension in this context can be interpreted as a vector quantity with values of correlation rising because of multiple definitions 
as various components. However, in principle when we define correlation in multi qubit situations, measurement in one subsystem can 
enhance the correlation in other two subsystems and thereby making quantum dissension to assume negative values \cite{Pati2012}. 
We emphasize on all possible one-particle projective measurements and two-particle projective measurements. 

\noindent The mutual information of three classical random variables in terms of entropies and joint entropies, are given by
\begin{multline}
I(X:Y:Z)=H(X)+H(Y)+H(Z) \\
-[H(X,Y)+H(X,Z)+H(Y,Z)]+H(X,Y,Z).
\end{multline}
\noindent It is also possible to obtain an expression for mutual information $I(X:Y:Z)$ that involves conditional entropy with respect to one random variable:
\begin{multline}
J(X:Y:Z)=H(X,Y)-H(X|Y)-H(Y|X) \\
-H(X|Z)-H(Y|Z)+H(X,Y|Z).
\end{multline}
\noindent One can define another equivalent expression for classical mutual information that includes conditional entropy with respect to two random variables:
\begin{multline}
K(X:Y:Z)=[H(X)+H(Y)+H(Z)] \\
-[H(X,Y)+H(X,Z)]+H(X|Y,Z).
\end{multline}
\noindent These equivalent classical information-theoretic definitions forms our basis for defining quantum dissension in the next subsections.

\subsection {Quantum Dissension for One-Particle Projective Measurement}
\noindent Let us consider a three-qubit state $\rho_{XYZ}$ where $X,Y$ and $Z$ refer to the first, second and the third qubit 
under consideration. The quantum version of $I(X:Y:Z)$ obtained by replacing random variables with density matrices and 
Shannon entropy with Von Neumann entropy reads,
\begin{multline}
I(X:Y:Z)=S(\rho_X)+S(\rho_Y)+S(\rho_Z) \\
-[S(\rho_{XY})+S(\rho_{YZ})+S(\rho_{XZ})]+S(\rho_{XYZ}).
\end{multline}
\noindent The quantum version of $J(X:Y:Z)$, obtained by appropriately defining conditional entropies, is given by
\begin{multline}
J(X:Y:Z)=S(\rho_{XY})-S(\rho_{Y|\Pi_{jX}})-S(\rho_{X|\Pi_{jY}}) \\
-S(\rho_{X|\Pi_{jZ}})-S(\rho_{Y|\Pi_{jZ}})+S(\rho^{X,Y|\Pi_{jZ}}),
\end{multline}
\noindent where $\Pi_j^n$ refer to a one particle projective measurement on the subsystem $'n'$ performed on the basis 
$\{|u_1\rangle$ = cos$\theta|0\rangle$ + $e^{i\phi}$ sin$\theta|1\rangle$,
  $|u_2\rangle$ = sin$\theta|0\rangle$ - $e^{i\phi}$ cos$\theta|1\rangle\}$
where $\theta$ and $\phi$ lies in the range [0,2$\pi$].

\noindent Quantum dissension function for single particle projective measurement is given by the difference of $I(X:Y:Z)$ and $J(X:Y:Z)$, i.e.
\begin{equation}
D_1(X:Y:Z)=J(X:Y:Z)-I(X:Y:Z).
\end{equation}
\noindent Quantum dissension is given by the quantity $\delta_1$ = min($D_1(X:Y:Z)$), where the minimization is taken over the entire range of basis parameters in order for $D_{1}$ to reveal maximum possible quantum correlation.

\subsection {Quantum Dissension for Two-Particle Projective Measurement}
\noindent The natural extension of $K(X:Y:Z)$ in the quantum domain is given by,
\begin{multline}
K(X:Y:Z)=[S(\rho_X)+S(\rho_Y)+S(\rho_Z)] \\
-[S(\rho_{XY})+S(\rho_{XZ})]+S(\rho_{X|{\Pi_{jYZ}}}).
\end{multline}

\noindent The two-particle projective measurement is carried out in the most general basis: 
$|v_1\rangle$ = cos$\theta|00\rangle$ + $e^{i\phi}$ sin$\theta|11\rangle$,
$|v_2\rangle$ = sin$\theta|00\rangle$ - $e^{i\phi}$ cos$\theta|11\rangle$,
$|v_3\rangle$ = cos$\theta|01\rangle$ + $e^{i\phi}$ sin$\theta|10\rangle$,
$|v_4\rangle$ = sin$\theta|01\rangle$ - $e^{i\phi}$ cos$\theta|10\rangle$,
where $\theta$, $\phi$ $\epsilon$ [0,2$\pi$]. 
In this case, the average quantum conditional entropy is given as $S(\rho_{X|YZ}) =\sum_jp_jS(\rho_{X|\Pi_{jYZ}})$ with
$p_{j}=tr[(I_X\otimes\Pi_{jYZ})\rho(I^X\otimes\Pi_{jYZ})]$ and $\rho^{X|\Pi_{jYZ}}=\frac{1}{p_j} tr_{YZ}[(I_X\otimes\Pi_{jYZ})\rho(I_X\otimes\Pi_{jYZ})]$. 

\noindent To define quantum dissension for two-particle projective measurement, we once again take the difference of the equivalent expressions of mutual information, i.e.
\begin{multline}
D_2(X:Y:Z)=K(X:Y:Z)-I(X:Y:Z) \\
=S(\rho_{X|\Pi_{jYZ}})+S(\rho_{YZ})-S(\rho_{XYZ}).
\end{multline}

\noindent The discord function $D_2$ is also interpreted as quantum discord with a bipartite split of the system. 
One can minimize $D_2$ over all two-particle measurement projectors to obtain dissension 
as $\delta_2$=min($D_2(X:Y:Z)$). This is the most generic expression since it includes all possible two-particle 
projective measurements. Both $\delta_1$ and $\delta_2$ together form the components of correlation vector defined 
in context of projective measurement done on different subsystems. 

\subsection {Monogamy of Quantum Correlations}
\noindent Monogamy of quantum correlation is an unique phenomenon which addresses distributed correlation in a multiparty setting. It states that in a multipartite situation, the total amount of individual correlations of a single party with other parties is bounded by the amount of correlation shared by the same party with the rest of the system when the rest are considered as a single entity. Mathematically, given a multipartite quantum state $\rho_{12...N}$ shared between N parties, the monogamy condition for a bipartite correlation measure $Q$ should satisfy $Q(\rho_{12})+Q(\rho_{13})+...+Q(\rho_{1N})$ $\leq$ $Q(\rho_{12...N})$ where $\rho_{1j}$ = $tr_{1...(j-1)(j+1)...N}$ $\rho_{1...j...N}$. It had been shown that certain entanglement measures satisfy the monogamy inequality . However, there are certain measures of quantum correlation, including quantum discord, which behave differently as far as the satisfying of monogamy inequality is concerned. By the term 'violation of monogamy inequality for certain measure', we actually refer to a situation where we can indeed find entangled states which violates the inequality for that measure. In case of quantum discord, it had been seen that W states violates the inequality and are polygamous in nature. 
More specifically, researchers considered the monogamy score $\delta_{m}$= $D(\rho_{AB})$+$D(\rho_{AC})$-$D(\rho_{A:BC})$, 
(where $\rho_{AB}$ and $\rho_{AC}$ are the traced out density matrices from $\rho_{ABC}$ and $D$ is 
quantum discord) and checked whether three-qubit states violate or satisfy 
the inequality $\delta_{m}$ $\leq$ 0.

\section {Effect of Noisy Channels on Quantum Dissension \& Monogamy Score}
\noindent In this section, we investigate the dynamics of quantum dissension when three-qubit states are transferred 
through noisy quantum channels. Moreover, we also study the change of the monogamy score for various initial states 
with time and purity of the state. We consider initial states to be mixed GHZ, mixed W, classical mixture of two 
separable states, a mixed biseparable states and the quantum channels to 
be amplitude damping, phase damping and depolarizing.

\noindent Given an initial state for three qubits $\rho(0)$, its evolution in the presence of quantum noise can be compactly written as,
\begin{equation}
\rho(t)=\sum_{l,m,n}K_{l,m,n}\rho(0){K^{\dag}}_{l,m,n},
\end{equation}
where $K_{l,m,n}$ are the Kraus operators satisfying $\sum_{l,m,n}K^{\dag}_{l,m,n}K_{l,m,n}$=I for all $t$ \cite{Nielsen2000,Kraus1983}. 
For independent channels, $K_{l,m,n}=K_{l}\otimes K_{m}\otimes K_{n}$ where $K_{\{l\}}$ describes one-qubit 
quantum channel effects.
We analytically present the dynamics of each initial state with respect to the individual channels. In other words we 
present the dynamics of each of $\delta_{1}$, $\delta_{2}$ and $\delta_{m}$. In each case, we apply the channel for sufficient time i.e. t=10 seconds.

\subsection{Effect of Generalized Amplitude Damping Channel}
\noindent In this subsection, we consider the effect of generalized amplitude damping channel on various three-qubit 
quantum states. The amplitude damping channel describes the process of energy dissipation in quantum processes such 
as spontaneous emission, spin relaxation, photon scattering and attenuation etc. It is described by single-qubit Kraus operators 
$K_{0}$=$\sqrt{q}$ diag(1,$\sqrt{1-\gamma}$), $K_{1}$=$\sqrt{q\gamma}$ ($\sigma_1+i\sigma_2$)/2, $K_{2}$=$\sqrt{1-q}$ 
diag($\sqrt{1-\gamma}$,1), $K_{3}$=$\sqrt{(1-q)\gamma}$ ($\sigma_1-i\sigma_2$)/2, where $q$ defines the final 
probability distribution when $T\rightarrow\infty$ (q=1 corresponds to the usual amplitude damping channel). Here 
$\gamma$=1-$e^{-\Gamma t}$, $\Gamma$ representing the decay rate.

\subsubsection{Dynamics of the channel for q=1}

\noindent\textit{\normalsize{1. Mixed GHZ State-}}\\

\noindent We consider the three-qubit mixed GHZ state $\rho_{GHZ}=(1-p)\frac{I}{8}+p|\textit{GHZ}\rangle\langle\textit{GHZ}|$ 
(we universally take $p$ as the classical randomness) as the initial state. The matrix elements of the density operator for a certain time $t$, or for a certain value of 
parameter $\gamma$ are given by,
\begin{eqnarray}
&& \rho_{11}=\frac{1}{8}(1+\gamma)[(1+\gamma)^2+3p(1-\gamma)^2],{}\nonumber\\&&
\rho_{22}=\rho_{33}=\rho_{55}=\frac{1}{8}(1-\gamma)[(1+\gamma)^2-p(1-\gamma)(3\gamma+1)],{}\nonumber\\&&
\rho_{44}=\rho_{66}=\rho_{77}=\frac{1}{8}(1-\gamma)^2[(1+\gamma)+p(3\gamma-1)],{}\nonumber\\&&
\rho_{88}=\frac{1}{8}(1+3p)(1-\gamma)^3,\rho_{18}=\frac{p}{2}(1-\gamma)^\frac{3}{2}.\
\end{eqnarray}
\noindent It is evident from Fig.1, $\delta_{1}$ and $\delta_{2}$ attains values $(-3.00,1.00)$ at $t=0$ and $p=1$ and decays asymptotically 
till each of them approaches $0$. The amplitude damping channel leaves the final 
population state at (diag$[1,0])^{\otimes 3}$ which contains no quantum dissension. In other words, we have a 
steady decay of quantum dissension for the mixed GHZ state with time in an amplitude damping channel.
The reduced density matrices are separable states and contain zero quantum discord 
for all values of time and purity. Thus, the monogamy score $\delta_{m}$ is negative of $\delta_{2}$. 
Since $\delta_{m}$ is always negative in this case, the state remains monogamous through out the evolution period.\\
\begin{figure}[h]
\begin{center}
\subfigure{\includegraphics[width=4cm,height=4cm]{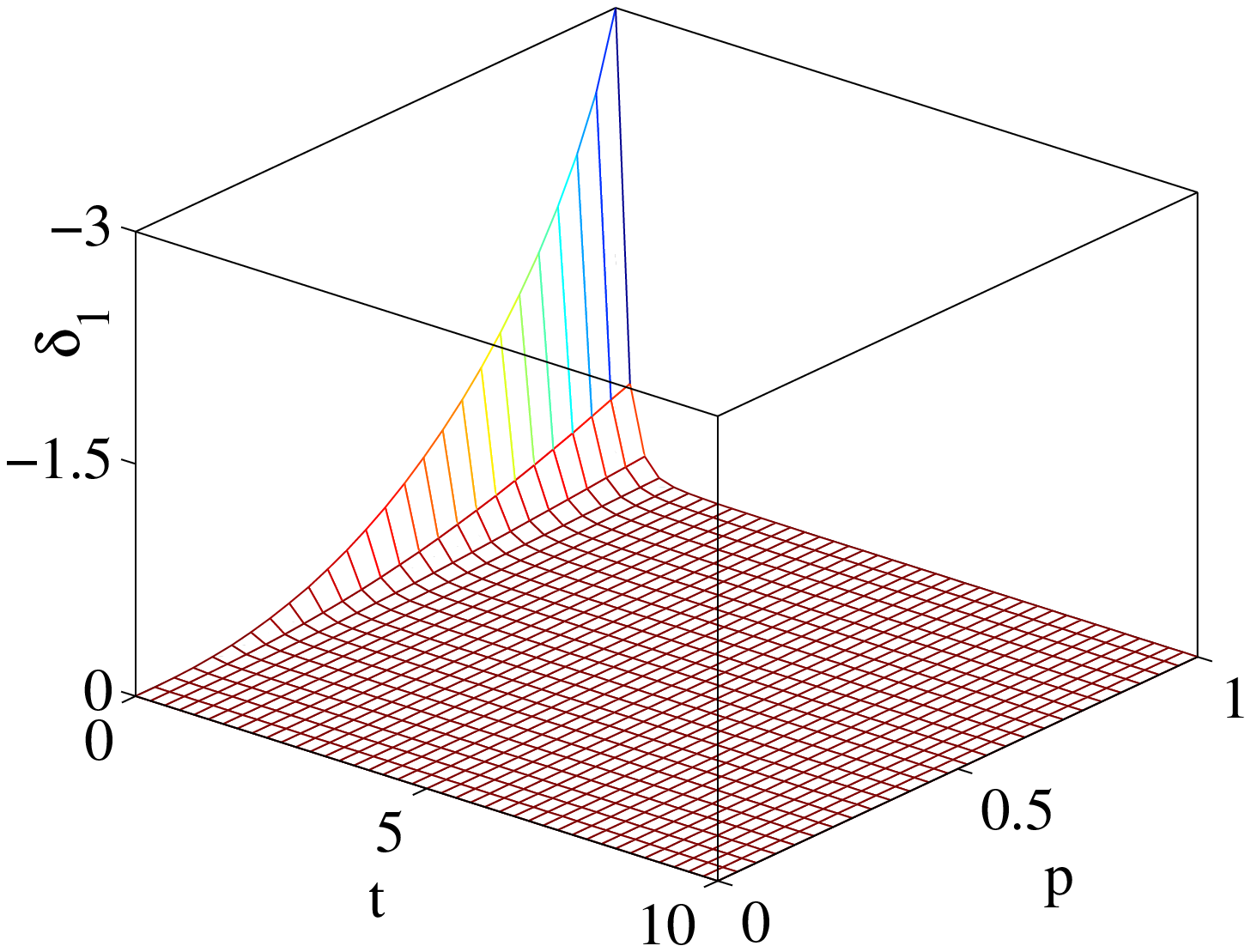}}
\subfigure{\includegraphics[width=4cm,height=4cm]{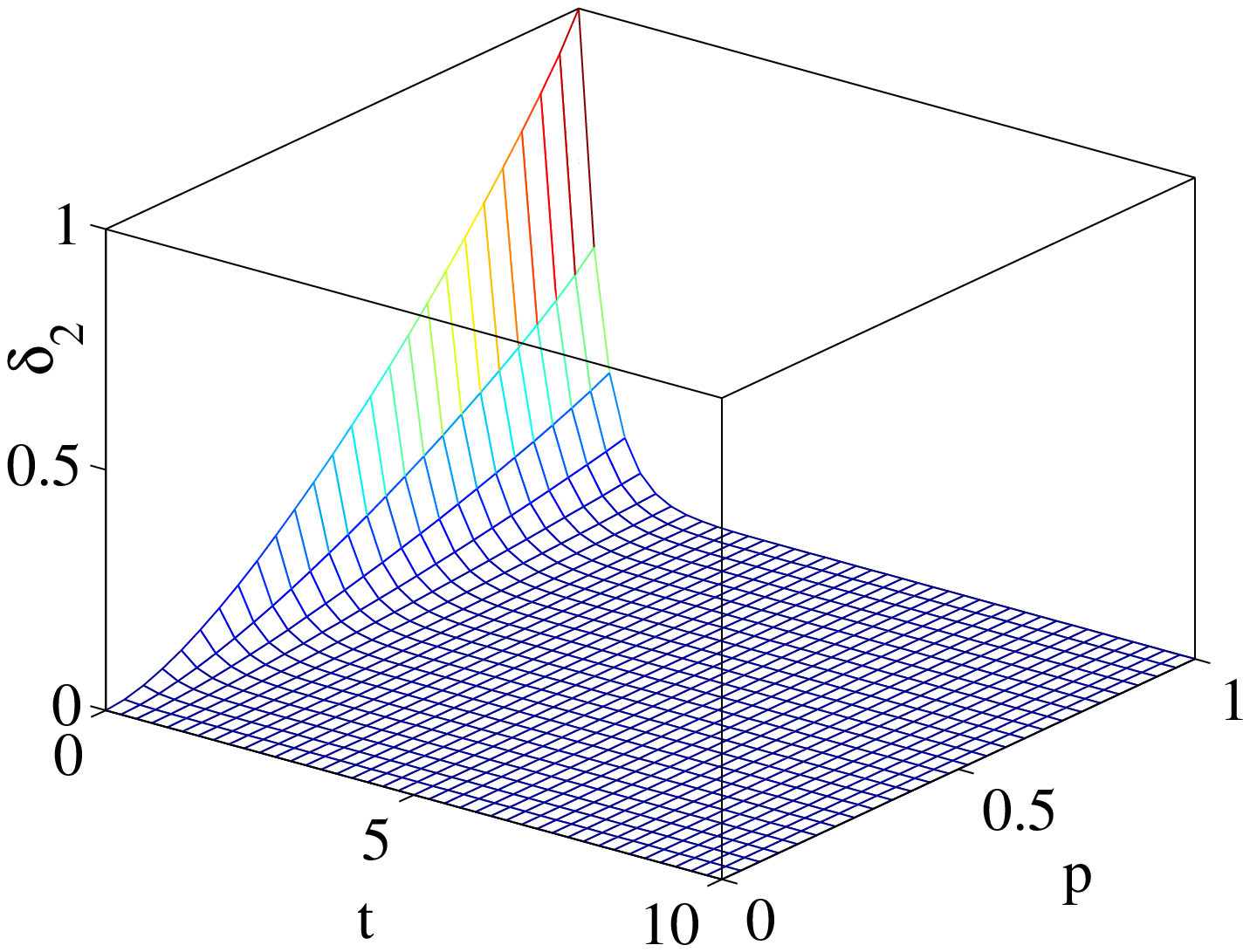}}
\caption{$\delta_{1}$ and $\delta_{2}$ dynamics of mixed GHZ state in GAD Channel with $q=1$}
\end{center}
\end{figure}

\noindent\textit{\normalsize{2. Mixed W State-}}\\

\noindent For the three-qubit mixed W 
state $\rho_{W}=(1-p)\frac{I}{8}+p|\textit{W}\rangle\langle\textit{W}|$, the dynamics of the state in terms of 
the matrix elements at time $t$ is given by,
\begin{eqnarray}
&&\rho_{11}=\frac{1}{8}[(1+\gamma)^3+p(1-\gamma)(\gamma^2+4\gamma-1)],{}\nonumber\\&&
\rho_{22}=\rho_{33}=\rho_{55}={}\nonumber\\&&
\frac{1}{24}(1-\gamma)[3(1+\gamma)^2-p(3\gamma^2+6\gamma-5)],{}\nonumber\\&&
\rho_{44}=\rho_{66}=\rho_{77}=\frac{1}{8}(1-p)(1-\gamma)^2 (1+\gamma),{}\nonumber\\&&
\rho_{88}=\frac{1}{8}(1-p)(1-\gamma)^3,{}\nonumber\\&&
\rho_{23}=\rho_{25}=\rho_{35}=\frac{1}{3}p(1-\gamma).
\end{eqnarray}
\noindent The initial values of $\delta_{1}$ and $\delta_{2}$ for a pure W state are (-1.75,0.92) respectively. 
As shown in Fig.2, $\delta_{1}$ and $\delta_{2}$ starts asymptotic decay from (-1.75,0.92) at $t=0$ and $p=1$ 
till they approach $0$ after sufficient channel action. The final population distribution at the 
limit of $\gamma\rightarrow$1 is (diag$[1,0])^{\otimes 3}$ resulting in zero quantum dissension. 
In Fig.3(a), we study the evolution of monogamy score with time and interestingly we find that for 
certain values of the parameter $p$, the monogamy score $\delta_{m}$ changes from negative to positive. 
This is a clear indication of the fact that the states which are initially monogamous are entering into 
the polygamous regime with time. 
\begin{figure}[h]
\begin{center}
\subfigure{\includegraphics[width=4cm,height=4cm]{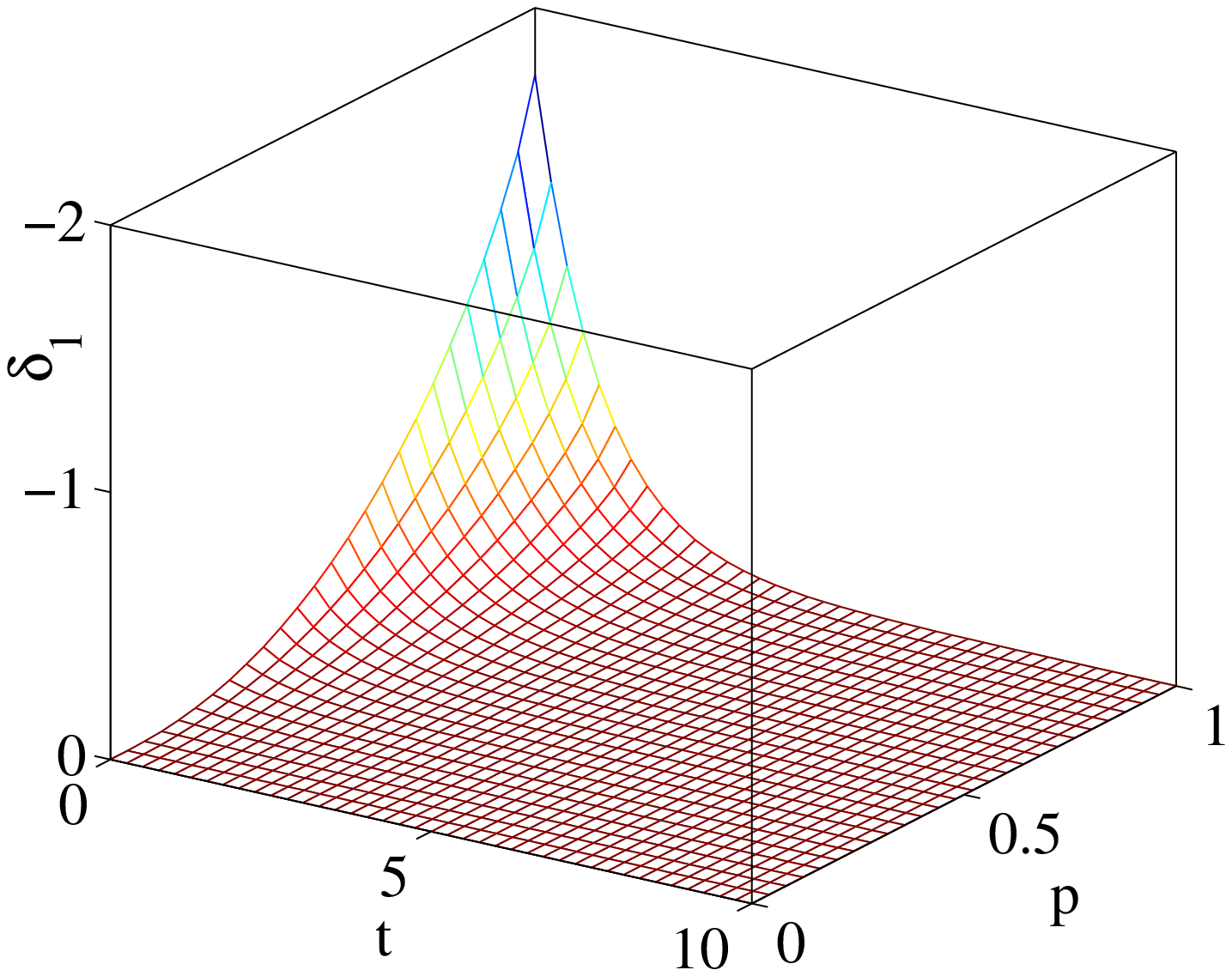}}
\subfigure{\includegraphics[width=4cm,height=4cm]{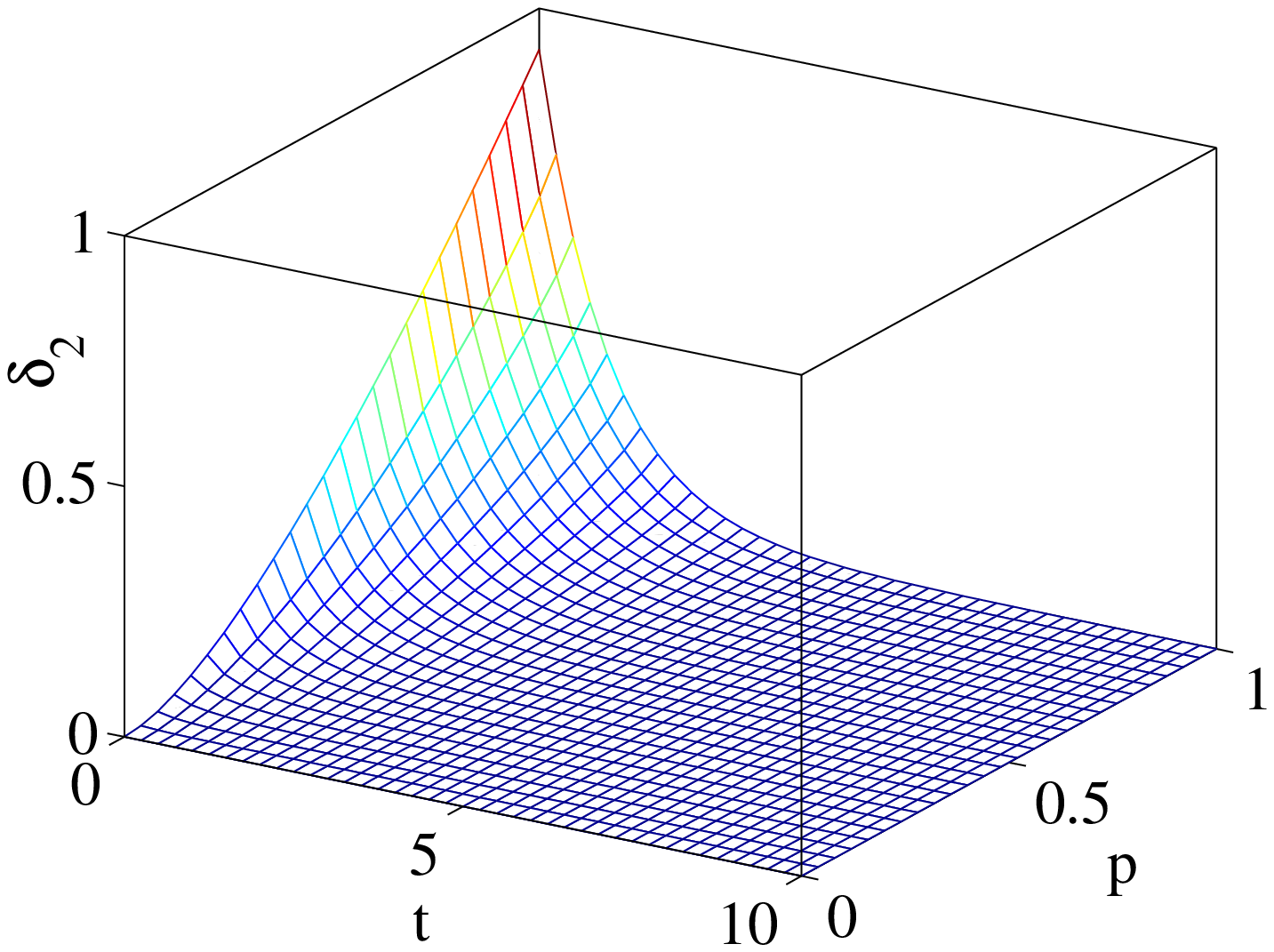}}
\caption{$\delta_{1}$ and $\delta_{2}$ dynamics of mixed W state in GAD Channel with $q=1$}
\end{center}
\end{figure}
\begin{figure}[h]
\begin{center}
\subfigure{\includegraphics[width=4cm,height=4cm]{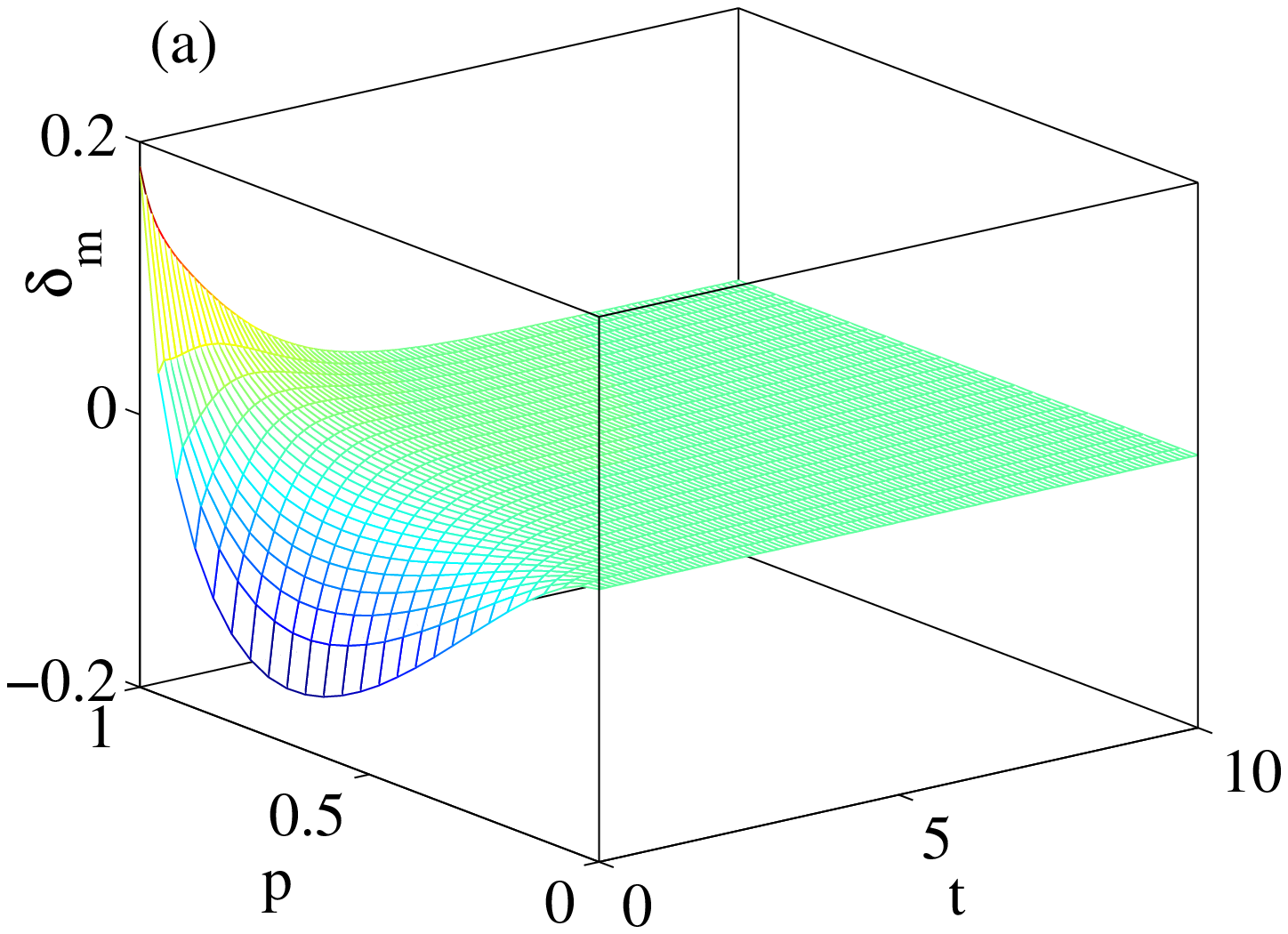}}
\subfigure{\includegraphics[width=4cm,height=4cm]{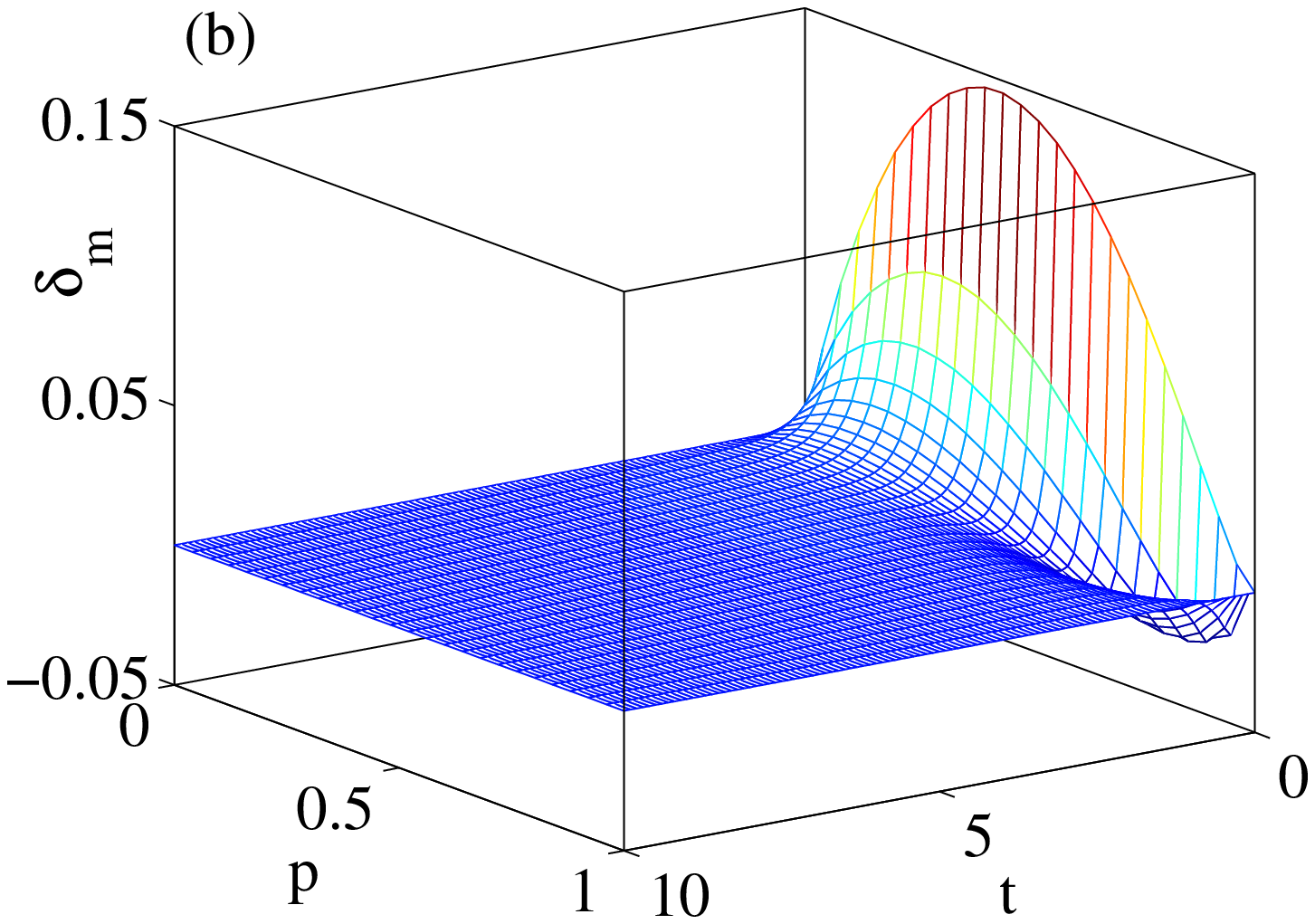}}
\caption{$\delta_{m}$ dynamics in GAD Channel with $q=1$ of (a) Mixed W state, (b) Mixture of separable states: $p$$|$000$\rangle$$\langle$000$|$+$(1-p)$$|$+++$\rangle$$\langle$+++$|$}
\end{center}
\end{figure}

\noindent \textit{\normalsize{3. Mixture of Separable States-}}\\

\noindent We take classical mixture of separable states $|000\rangle$ and $|+++\rangle$ given by the density 
matrix $\rho=p|000\rangle\langle000|+(1-p)|+++\rangle\langle +++|$. The dynamics of this 
mixture under the action of amplitude damping channel in terms of the matrix elements is as follows
\begin{eqnarray}
&&\rho_{11}=\frac{1}{8}[(1+\gamma)^3+p(1-\gamma)(\gamma^2+4\gamma+7)],{}\nonumber\\&&
\rho_{12}=\rho_{13}=\rho_{15}=\frac{1}{8}(1-p)\sqrt{1-\gamma}(1+\gamma)^2,{}\nonumber\\&&
\rho_{14}=\rho_{16}=\rho_{17}=\rho_{23}=\rho_{25}=\rho_{35}=\frac{1}{8}(1-p)(1-\gamma^2),{}\nonumber\\&&
\rho_{18}=\rho_{27}=\rho_{36}=\rho_{45}=\frac{1}{8}(1-p)(1-\gamma)^\frac{3}{2},{}\nonumber\\&&
\rho_{22}=\rho_{33}=\rho_{55}=\frac{1}{8}(1-p)(1-\gamma)(1+\gamma)^2,{}\nonumber\\&&
\rho_{24}=\rho_{26}=\rho_{34}=\rho_{37}=\rho_{56}=\rho_{57}={}\nonumber\\&&
\frac{1}{8}(1-p)\sqrt{1-\gamma}(1-\gamma^2),{}\nonumber\\&&
\rho_{44}=\rho_{66}=\rho_{77}=\frac{1}{8}(1-p)(1-\gamma)^2(1+\gamma),{}\nonumber\\&&
\rho_{28}=\rho_{38}=\rho_{46}=\rho_{47}=\rho_{58}=\rho_{67}=\frac{1}{8}(1-p)(1-\gamma)^2,{}\nonumber\\&&
\rho_{48}=\rho_{68}=\rho_{78}=\frac{1}{8}(1-p)(1-\gamma)^\frac{5}{2},{}\nonumber\\&&
\rho_{88}=\frac{1}{8}(1-p)(1-\gamma)^3.
\end{eqnarray}
At $t=0$, the maximum values (-1.015,0.15) of quantum dissensions $\delta_{1}$ and $\delta_{2}$ are obtained for $p=1/2$ [Fig.4].
In this particular dynamics, we observe an interesting phenomenon that there is no exact asymptotic decay of quantum dissension $\delta_{1}$. We observe the revival of quantum correlation for a certain period of time in the initial phase of the dynamics. This is something different from the standard intuition of asymptotic decay of quantum correlation when it undergoes dissipative dynamics. This remarkable feature can be interpreted as that the dissipative dynamics is not necessarily going to decrease quantum correlation with passage of time. On the contrary, depending upon the initial state it can enhance the quantum correlation for a certain period of time. We refer to this unique feature as \textit{revival of quantum correlation in dissipative dynamics}. However, the other dissension $\delta_{2}$ follows the standard process of asymptotic decay with time.

\noindent In Fig.3(b), we also compute the monogamy score and find that states which are 
initially polygamous are becoming monogamous with the passage of time. This is contrary to what we observed in 
case of mixed W states. In this case, the states initially freely shareable (polygamous) are entering into not 
freely shareable (monogamous) regime due to channel action. This is a remarkable feature as this helps us 
to obtain monogamous state 
from polygamous state. This is indeed helpful as monogamy of quantum correlation is an useful tool for quantum security.

\begin{figure}[h]
\begin{center}
\subfigure{\includegraphics[width=4cm,height=4cm]{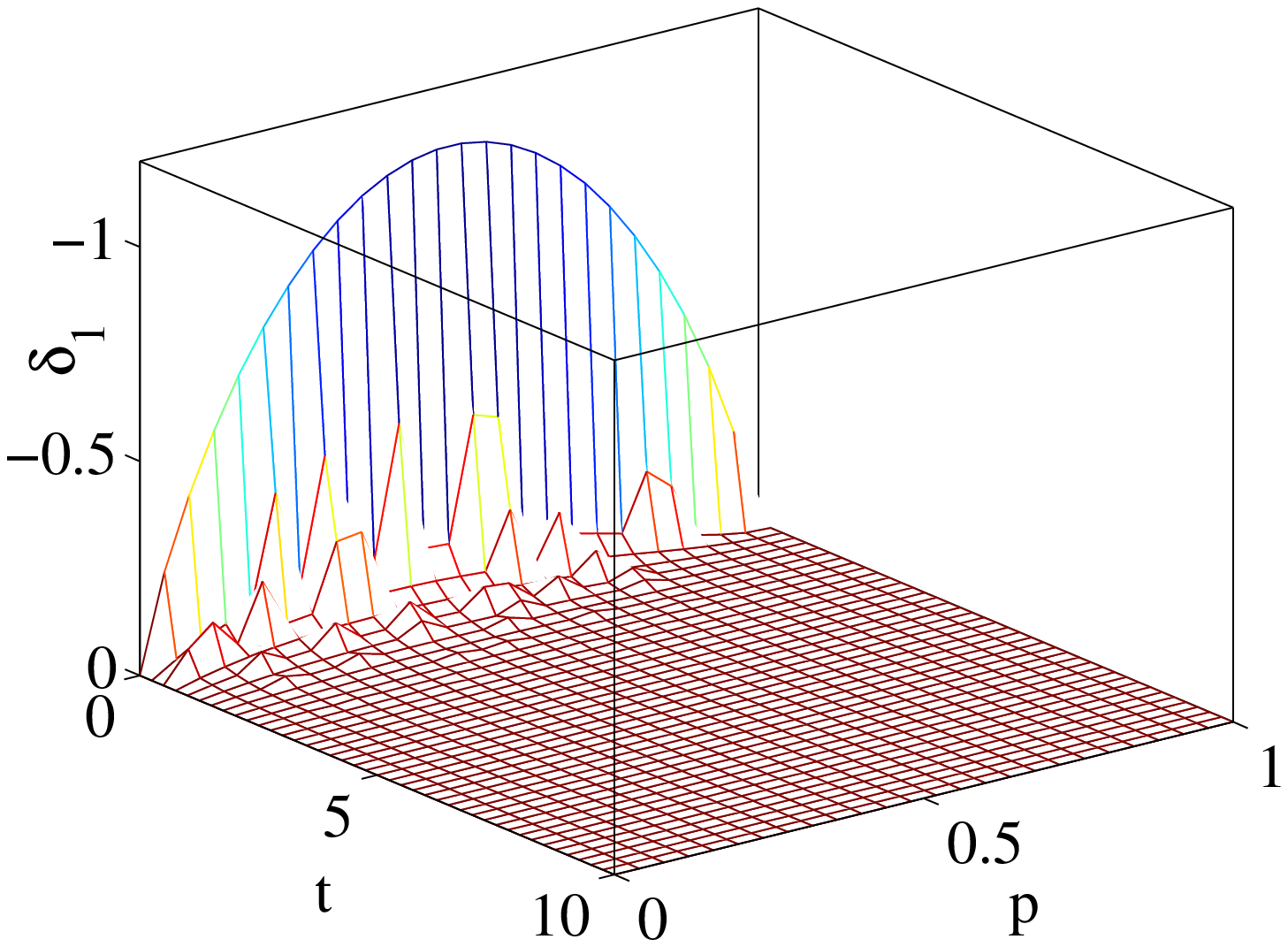}}
\subfigure{\includegraphics[width=4cm,height=4cm]{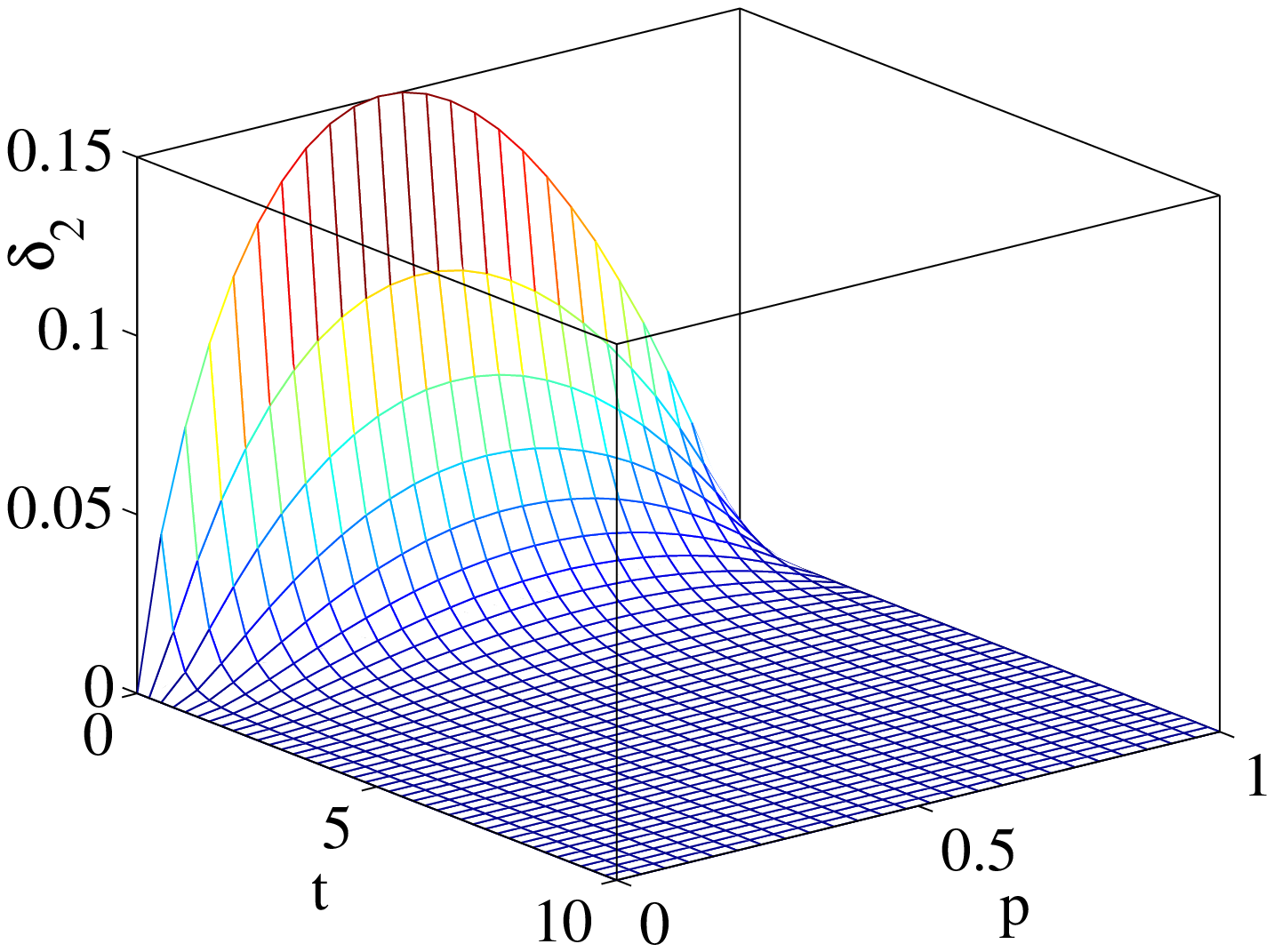}}
\caption{$\delta_{1}$ and $\delta_{2}$ dynamics of mixture of separable states $|000\rangle$ and $|+++\rangle$ in GAD Channel with $q=1$}
\end{center}
\end{figure}

\noindent \textit{\normalsize{4. Mixed Biseparable State-}}\\

\noindent Now we provide another example where action of quantum noisy channel can revive quantum dissension for a short period of time in a much smooth manner compared to our previous example. Here, we consider a mixed biseparable state: $\rho$=$(1-p)$$\frac{I}{8}$+$\frac{p}{2}$[$|0\rangle$$|\varphi^+\rangle$$\langle0|$$\langle\varphi^+|$+$|0\rangle$$|\psi^-\rangle$$\langle0|$$\langle\psi^-|$]. The dynamics of this density matrix at time $t$ is given by,
\begin{eqnarray}
&& \rho_{11}=\frac{1}{8}(1+\gamma)^2[(1+\gamma)+p(1-\gamma)],{}\nonumber\\&&
\rho_{22}=\rho_{33}=\frac{1}{8}(1-\gamma^2)[(1+\gamma)+p(1-\gamma)],{}\nonumber\\&&
\rho_{44}=\frac{1}{8}(1-\gamma)^2[(1+\gamma)+p(1-\gamma)],{}\nonumber\\&&
\rho_{55}=\frac{1}{8}(1-p)(1-\gamma)(1+\gamma)^2,{}\nonumber\\&&
\rho_{66}=\rho_{77}=\frac{1}{8}(1-p)(1-\gamma)^2(1+\gamma),{}\nonumber\\&&
\rho_{88}=\frac{1}{8}(1-p)(1-\gamma)^3,{}\nonumber\\&&
\rho_{14}=-\rho_{23}=\frac{p}{4}(1-\gamma).
\end{eqnarray}
\noindent At $t=0$ for this state, both $\delta_{1}$ and $\delta_{2}$ are having the value 0. However, quite surprisingly, 
we find that in the initial phase both dissension $\delta_{1}$ and $\delta_{2}$ increase and attain maximum 
values (0.00133,0.00183) and in the subsequent phases the values lower down and finally reach 0 [Fig.5]. 
This reiterates the fact that for certain initial states the dissipative dynamics acts as a catalyst and helps in 
revival of quantum correlation. This dynamics is different from our previous dynamics in the sense that here 
revival of quantum correlation is much more than the quantum correlation present in the initial state. This is 
indeed a strong signature that in multi-qubit cases the channel dynamics can take a zero-correlated to a correlated 
state. Though the rise of correlation is not very high, however, in NMR systems \cite{Mahesh2011} this rise is 
significant as one starting with a zero-correlated state can use the state for computation at subsequent phases of 
time instead of trashing it away. The reduced density matrices  are separable states for 
all values of time and purity, making making their discord equal to zero. Here once again we have, 
$\delta_{m}$ = -$\delta_{2}$ and hence channel action does not change the monogamy property of the mixed biseparable state.
\begin{figure}[h]
\begin{center}
\subfigure{\includegraphics[width=4cm,height=4cm]{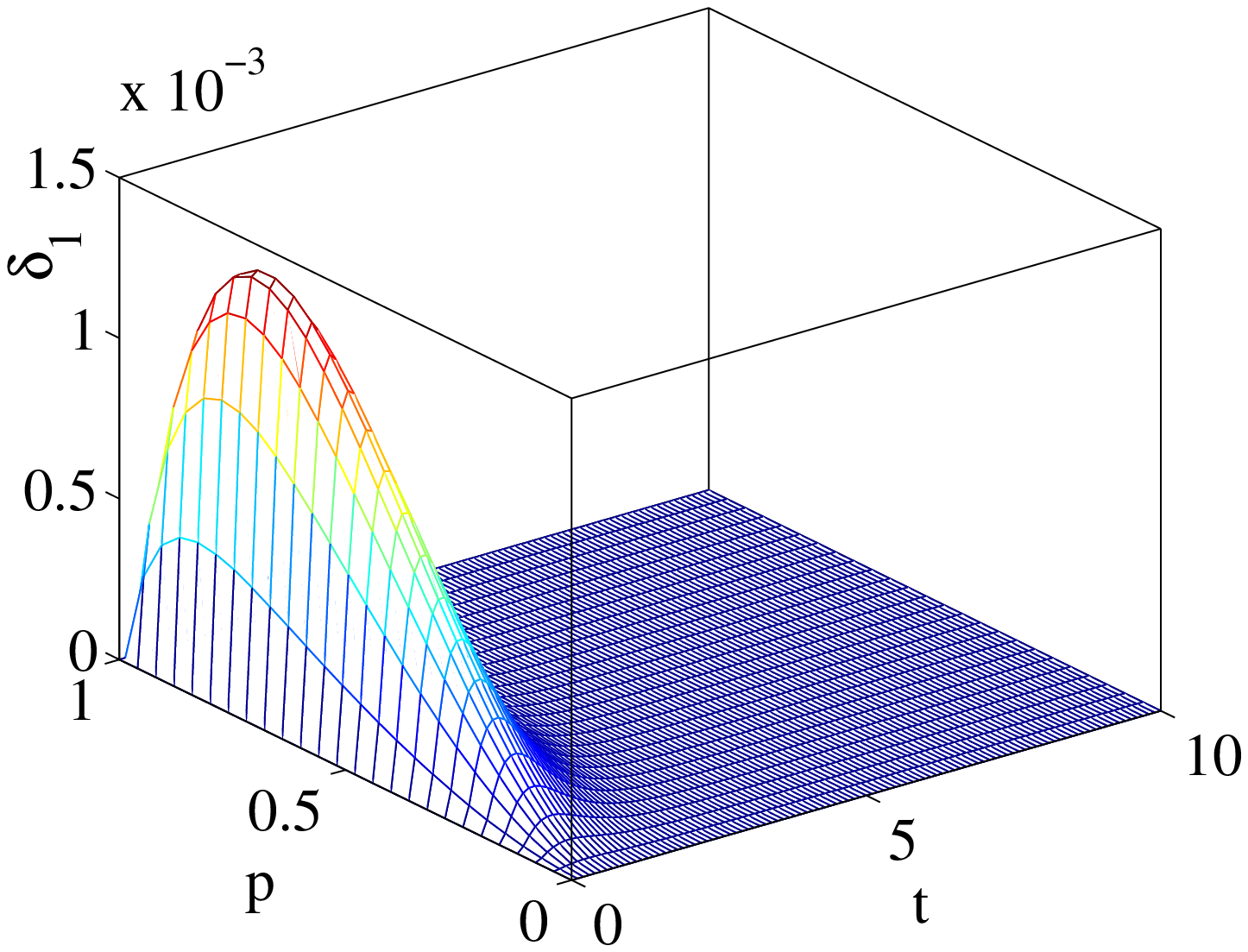}}
\subfigure{\includegraphics[width=4cm,height=4cm]{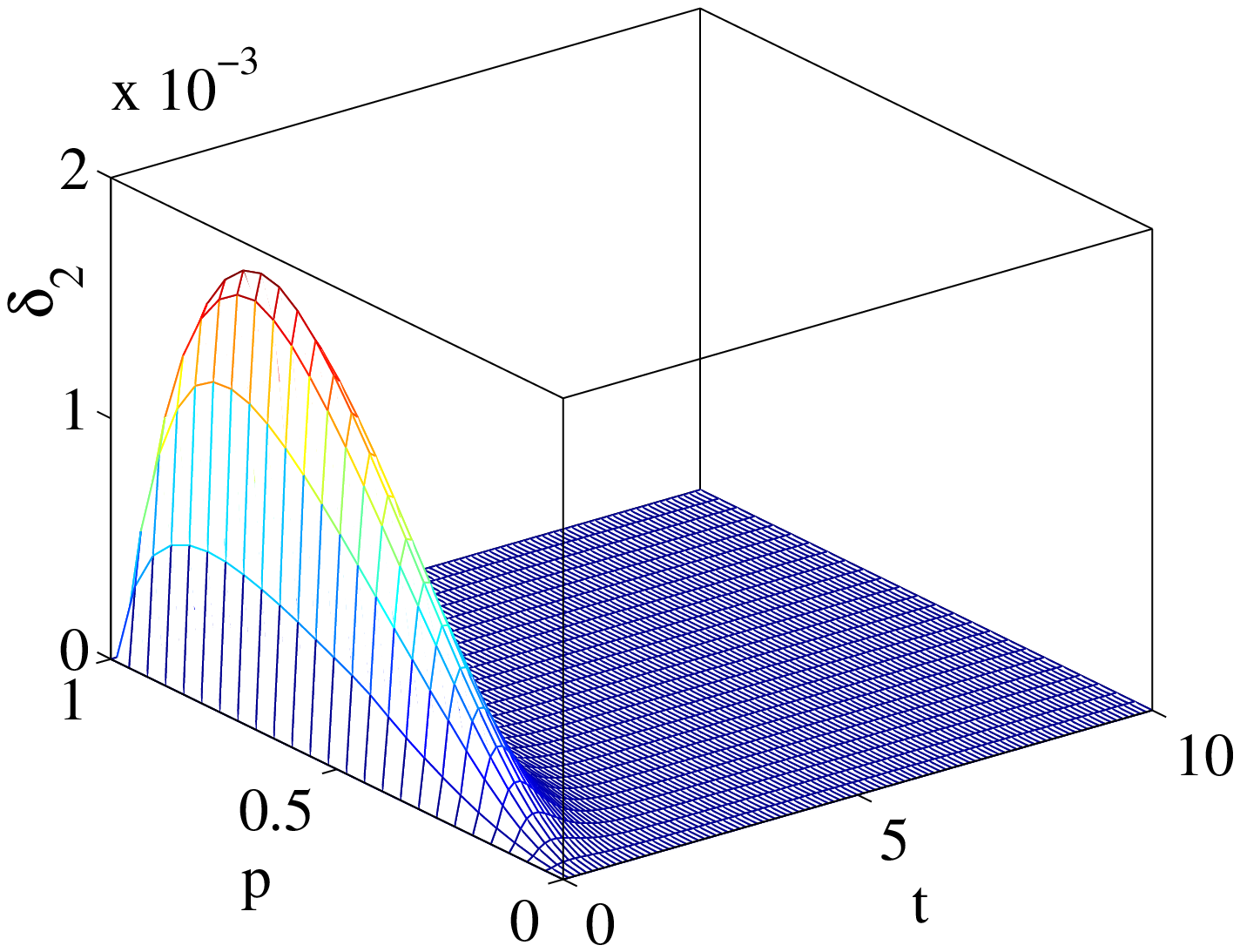}}
\caption{$\delta_{1}$ and $\delta_{2}$ dynamics of mixed biseparable state in GAD channel with $q=1$.}
\end{center}
\end{figure}

\subsubsection{Dynamics of the channel for q=1/2}

\noindent \textit{\normalsize{1. Mixed GHZ State-}}\\

\noindent The density matrix elements of mixed GHZ at time $t$ for $q=1/2$ are given as,
\begin{eqnarray}
&&\rho_{11}=\rho_{88}=\frac{1}{8}[1+3p(1-\gamma)^2],\rho_{18}=\frac{p}{2}(1-\gamma)^\frac{3}{2},{}\nonumber\\&&
\rho_{ii}=\frac{1}{8}[1-p(1-\gamma)^2] , i=2,...,7.\\
\end{eqnarray}
Here $\delta_{1}$ and $\delta_{2}$ starts decaying from (-3.00,1.00) at $t=0$, $p=1$ and approaches $0$ after 
sufficient time [Fig.6]. The decay of $\delta_{1}$ is not exactly asymptotic in contrast to the action of 
GAD channel with q=1. The decay of $\delta_{2}$ is asymptotic as in the case of GAD channel with $q=1$. The 
initial state evolves to final population distribution (diag$[1/2,1/2])^{\otimes 3}$, which contains no quantum 
dissension. Moreover, the reduced density matrices  are separable states and do not 
contribute towards monogamy score. Therefore, dynamics of $\delta_{m}$ is same as that of $\delta_{2}$, only 
differing by a negative sign.\\
  
\begin{figure}[h]
\begin{center}
\subfigure{\includegraphics[width=4cm,height=4cm]{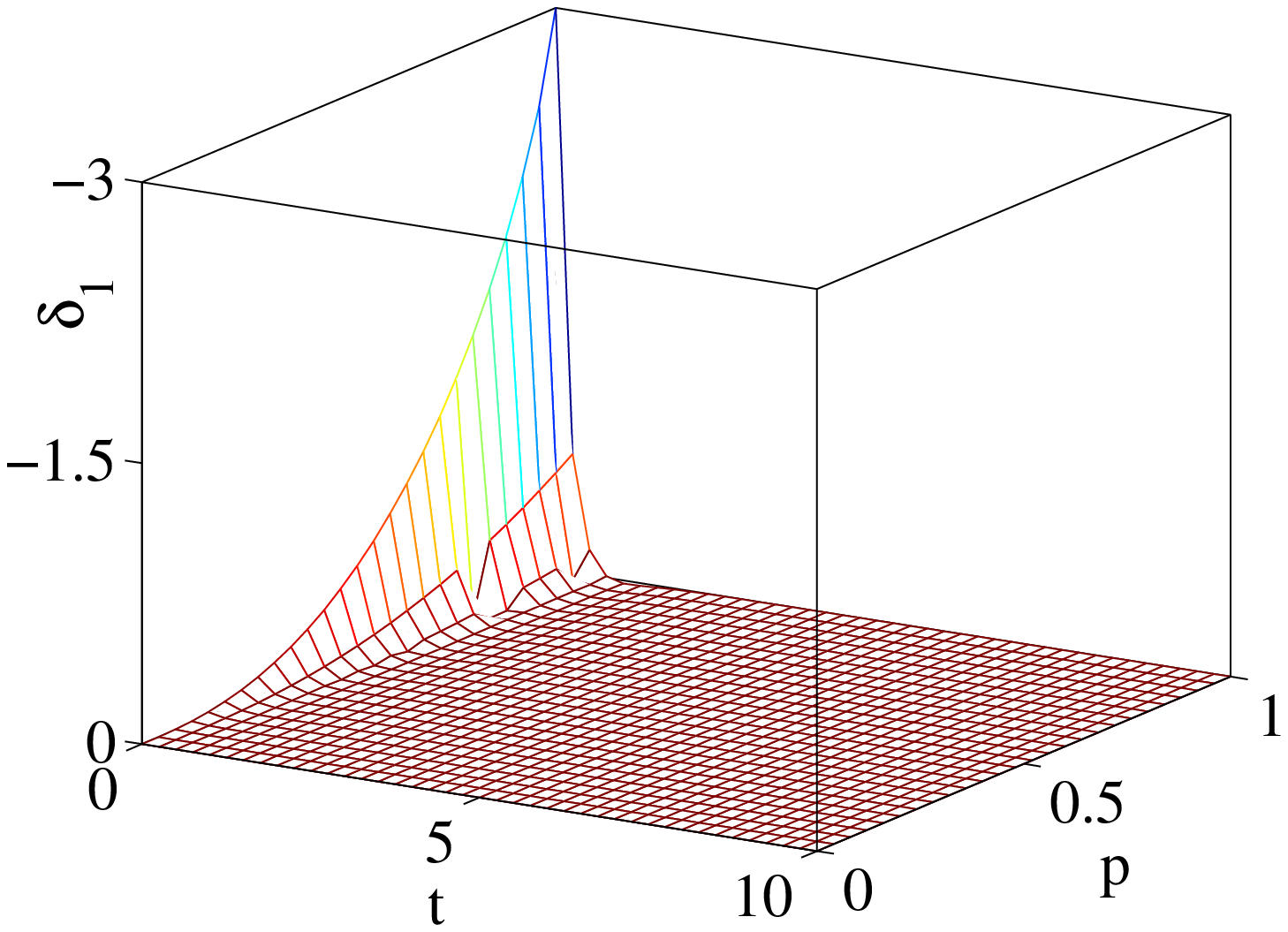}}
\subfigure{\includegraphics[width=4cm,height=4cm]{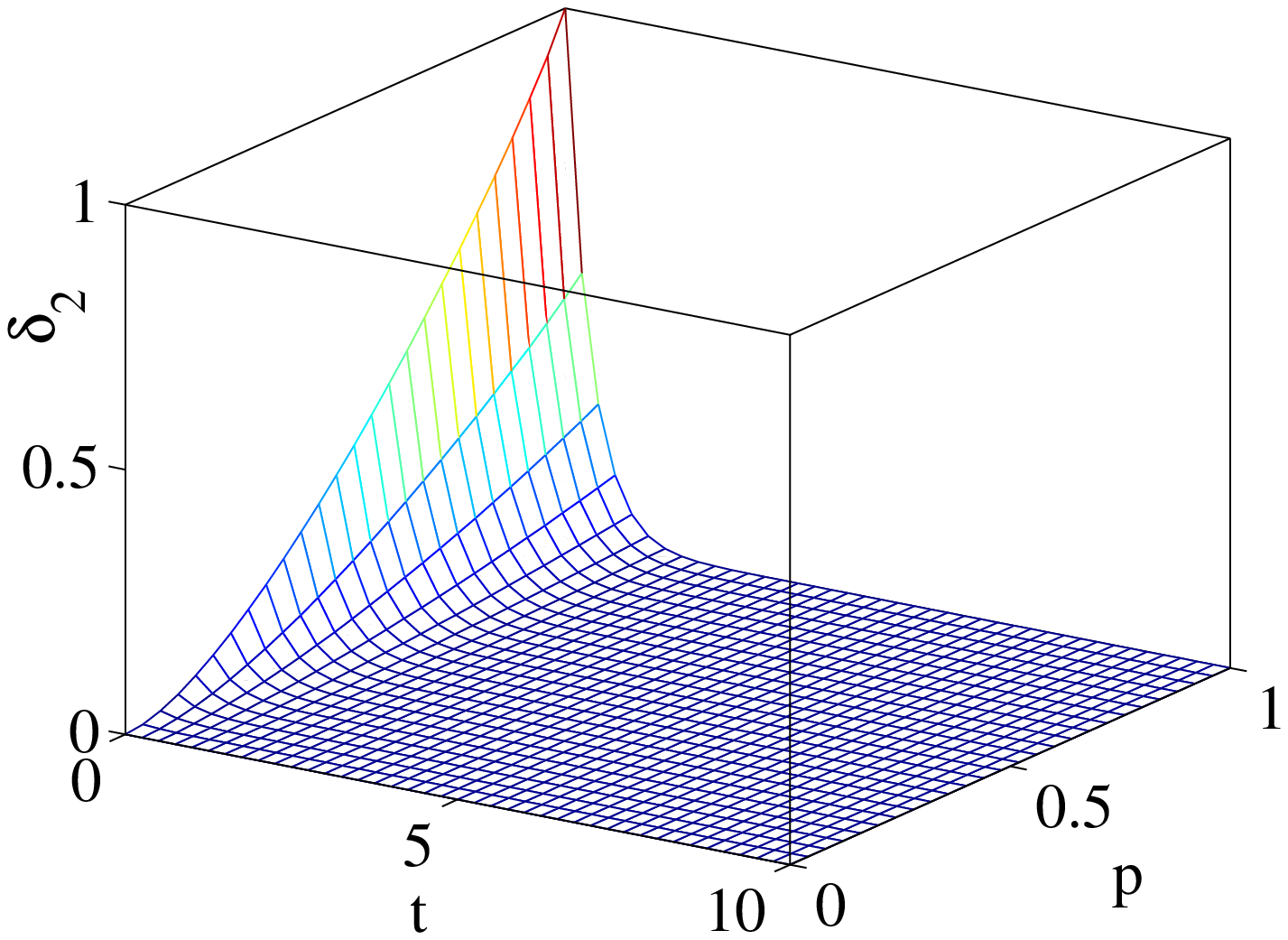}}
\caption{$\delta_{1}$ and $\delta_{2}$ dynamics of mixed GHZ state in GAD Channel with $q=1/2$}
\end{center}
\end{figure}

\noindent \textit{\normalsize{2. Mixed W State-}}\\

\noindent The density matrix evolution of the mixed W state at time $t$ is given by,
\begin{eqnarray}
&&\rho_{11}=\frac{1}{8}[1-p(1-\gamma)(\gamma^2-3\gamma+1)],{}\nonumber\\&&
\rho_{22}=\rho_{33}=\rho_{55}=\frac{1}{24}[3+p(1-\gamma)(3\gamma^2-7\gamma+5)],{}\nonumber\\&&
\rho_{44}=\rho_{66}=\rho_{77}=\frac{1}{24}[3-p(1-\gamma)(3\gamma^2-5\gamma+3)],{}\nonumber\\&&
\rho_{88}=\frac{1}{8}[1+p(1-\gamma)(\gamma^2-\gamma-1)],{}\nonumber\\&&
\rho_{23}=\rho_{25}=\rho_{35}=\frac{p}{6}(1-\gamma)(2-\gamma),{}\nonumber\\&&
\rho_{46}=\rho_{47}=\rho_{67}=\frac{p}{6}\gamma(1-\gamma).{}\nonumber\\&&
\end{eqnarray}
The quantum dissensions $\delta_{1}$ and $\delta_{2}$ attain values of (-1.75,0.92) at $t=0$ and $p=1$ [Fig.7]. 
As in the case of mixed GHZ state, the curve for $\delta_{1}$ is not exactly asymptotic while the curve 
for $\delta_{2}$ is asymptotic. In the limit $\gamma\rightarrow 1$, a final population distribution 
of (diag$[1/2,1/2])^{\otimes 3}$ is left resulting in zero quantum dissension. For purity values closer to 1, 
the initial states are polygamous and they enter into the monogamy regime due to action of GAD channel [Fig.8(a)]. 
The states with purity values closer to 0 are monogamous and do not experience any such transition. Hence once again we 
have one such example where there is a useful transition from polygamous to monogamous regime. 
\begin{figure}[h]
\begin{center}
\subfigure{\includegraphics[width=4cm,height=4cm]{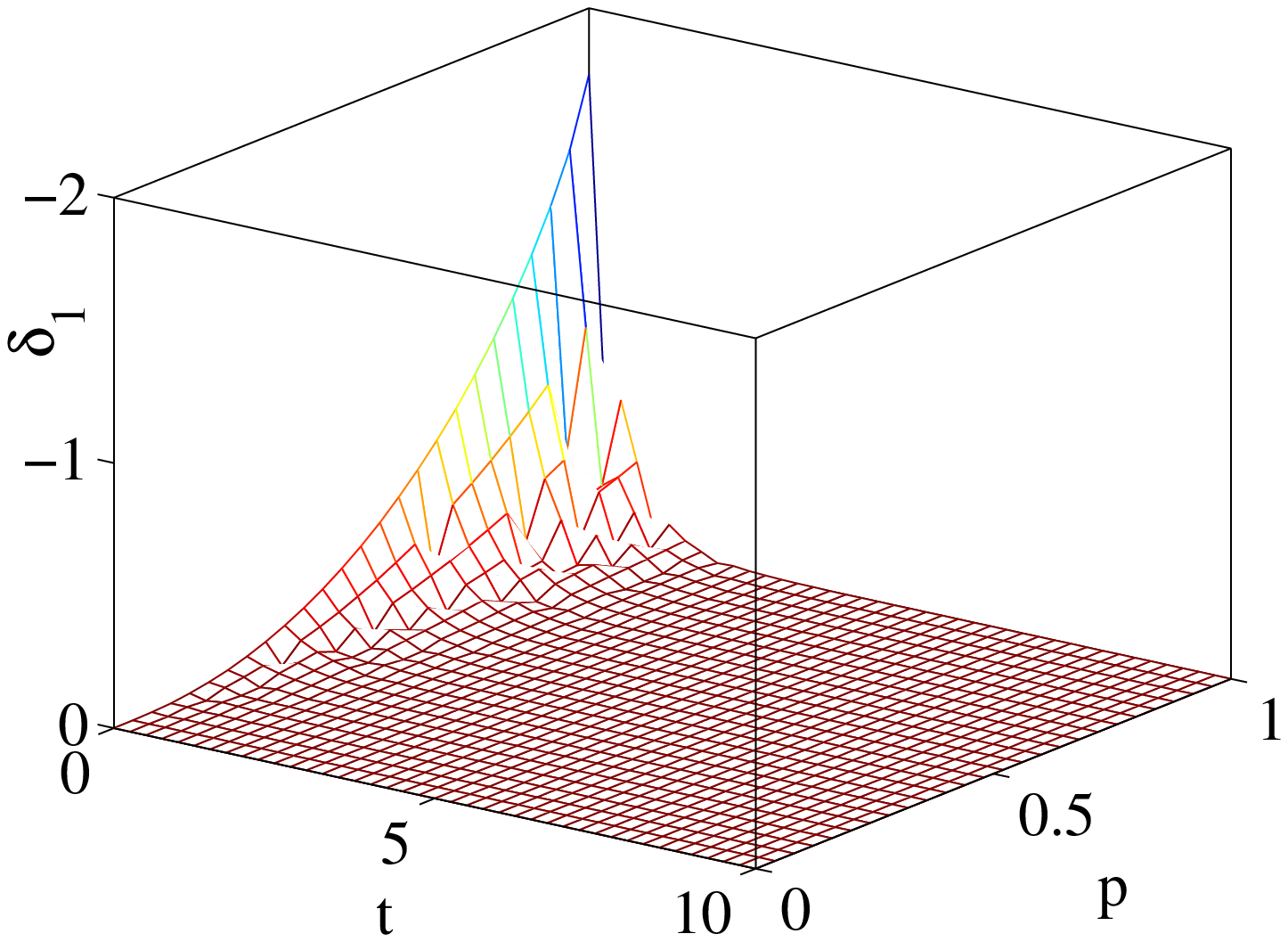}}
\subfigure{\includegraphics[width=4cm,height=4cm]{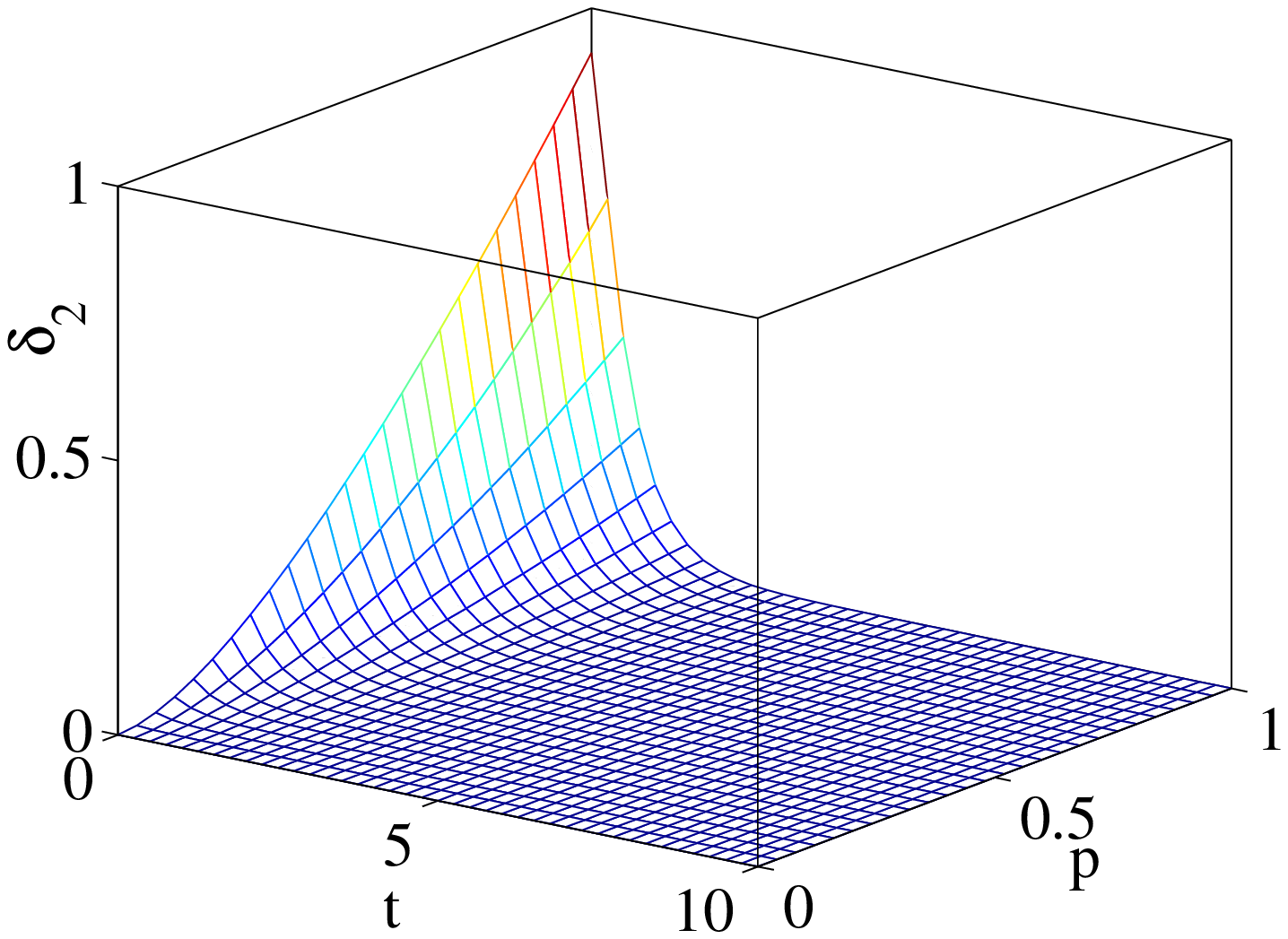}}
\caption{$\delta_{1}$ and $\delta_{2}$ dynamics of mixed W state in GAD Channel with $q=1/2$}
\end{center}
\end{figure}

\begin{figure}[h]
\begin{center}
\subfigure{\includegraphics[width=4cm,height=4cm]{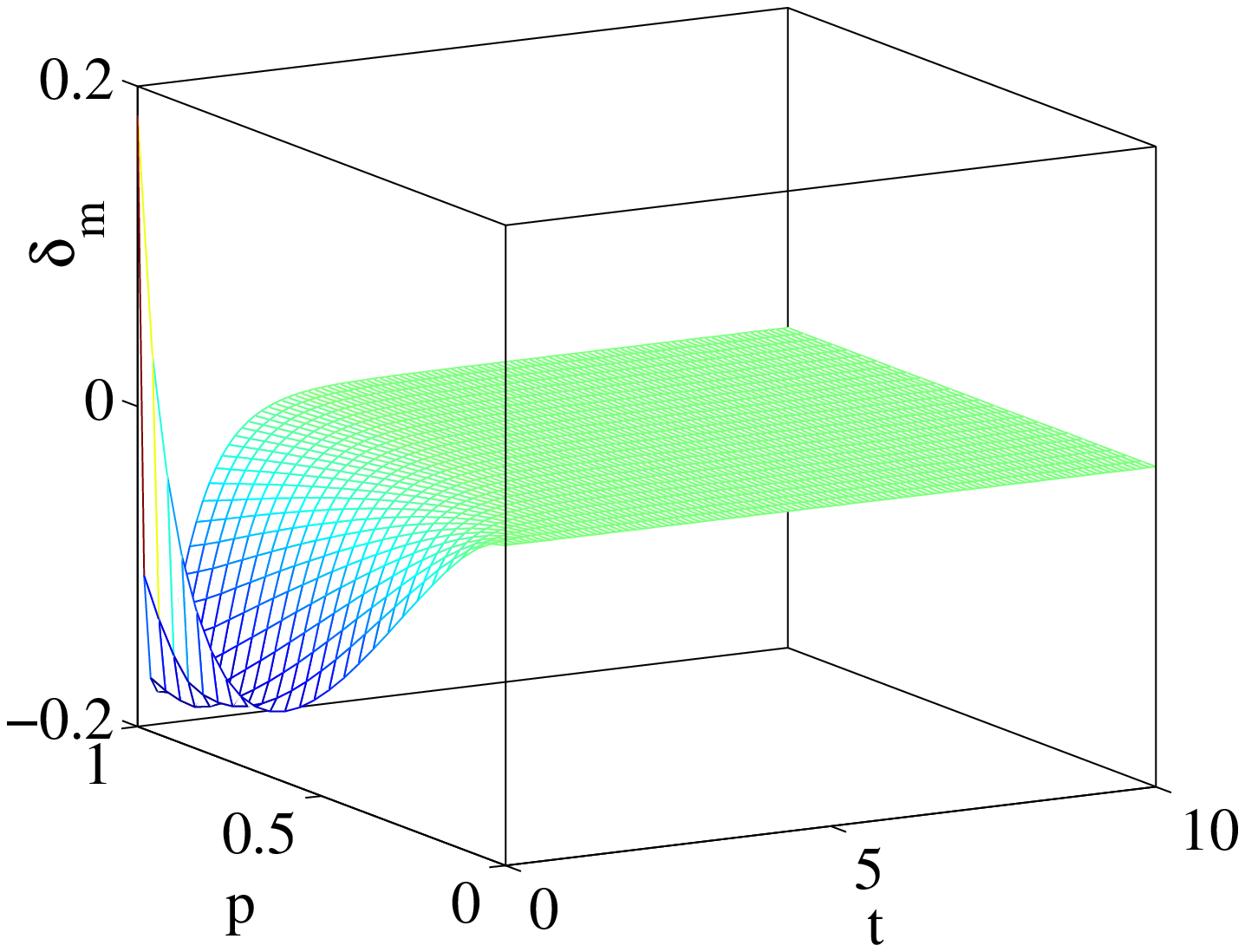}}
\subfigure{\includegraphics[width=4cm,height=4cm]{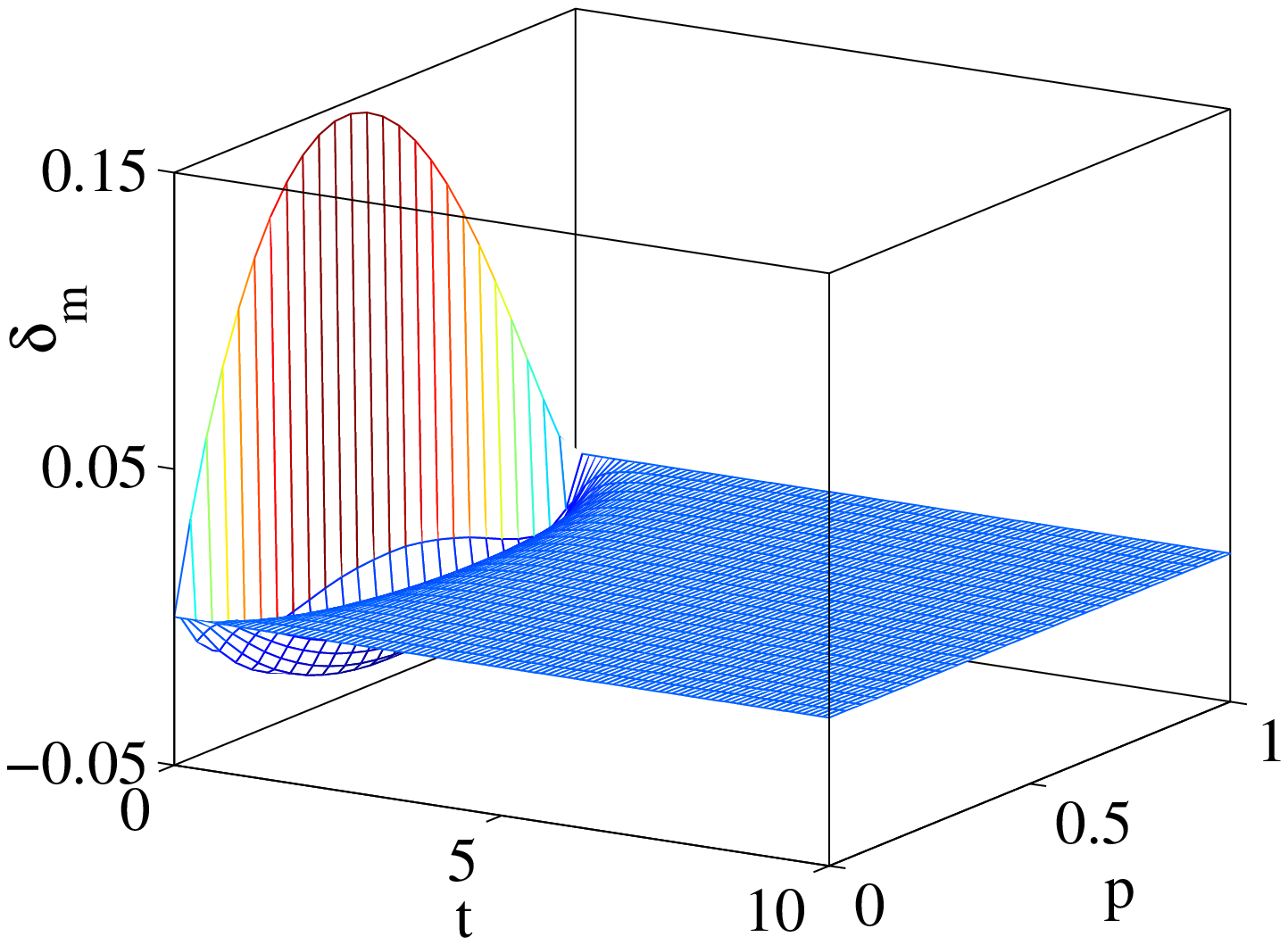}}
\caption{$\delta_{m}$ dynamics in GAD Channel with $q=1/2$ of (a) Mixed W state (b) Mixture of separable states: $p$$|$000$\rangle$$\langle$000$|$+$(1-p)$$|$+++$\rangle$$\langle$+++$|$}
\end{center}
\end{figure}

\noindent \textit{\normalsize{3. Mixture of Separable States-}}\\

\noindent We consider initial density matrix $\rho=p|000\rangle+(1-p)|+++\rangle$ 
whose dynamics at time $t$ is as follows,
\begin{eqnarray}
&&\rho_{11}=\frac{1}{8}[1+p(1-\gamma)(\gamma^2-5\gamma+7)],{}\nonumber\\&&
\rho_{22}=\rho_{33}=\rho_{55}=\frac{1}{8}[1-p(1-\gamma)(\gamma^2-3\gamma+1)],{}\nonumber\\&&
\rho_{44}=\rho_{66}=\rho_{77}=\frac{1}{8}[1+p(1-\gamma)(\gamma^2-\gamma-1)],{}\nonumber\\&&
\rho_{88}=\frac{1}{8}[1+p(\gamma^3-1)],{}\nonumber\\&&
\rho_{12}=\rho_{13}=\rho_{15}=\rho_{24}=\rho_{26}=\rho_{34}=\rho_{37}=\rho_{48}=\rho_{56}{}\nonumber\\&&
=\rho_{57}=\rho_{68}=\rho_{78}=\frac{1}{8}(1-p)\sqrt{1-\gamma},{}\nonumber\\&&
\rho_{14}=\rho_{16}=\rho_{17}=\rho_{23}=\rho_{25}=\rho_{28}=\rho_{35}=\rho_{38}=\rho_{46}{}\nonumber\\&&
=\rho_{47}=\rho_{58}=\rho_{67}=\frac{1}{8}(1-p)(1-\gamma),{}\nonumber\\&&
\rho_{18}=\rho_{27}=\rho_{36}=\rho_{45}=\frac{1}{8}(1-p)(1-\gamma)^\frac{3}{2}.
\end{eqnarray}

\noindent Once again it is evident from Fig[9], $\delta_{1}$ and $\delta_{2}$ achieve maximum values (-1.015,0.15) 
at $t=0$ and $p=1/2$. However, the decay profile of $\delta_{2}$ is much smoother than that of $\delta_{1}$. 
The evolution of monogamy score [Fig. 8(b)] is quite different for 
$q=1/2$ than that of $q=1$. Here also, all the initial polygamous density matrices enter into the 
monogamy regime irrespective of the values of parameter p.
\begin{figure}[h]
\begin{center}
\subfigure{\includegraphics[width=4cm,height=4cm]{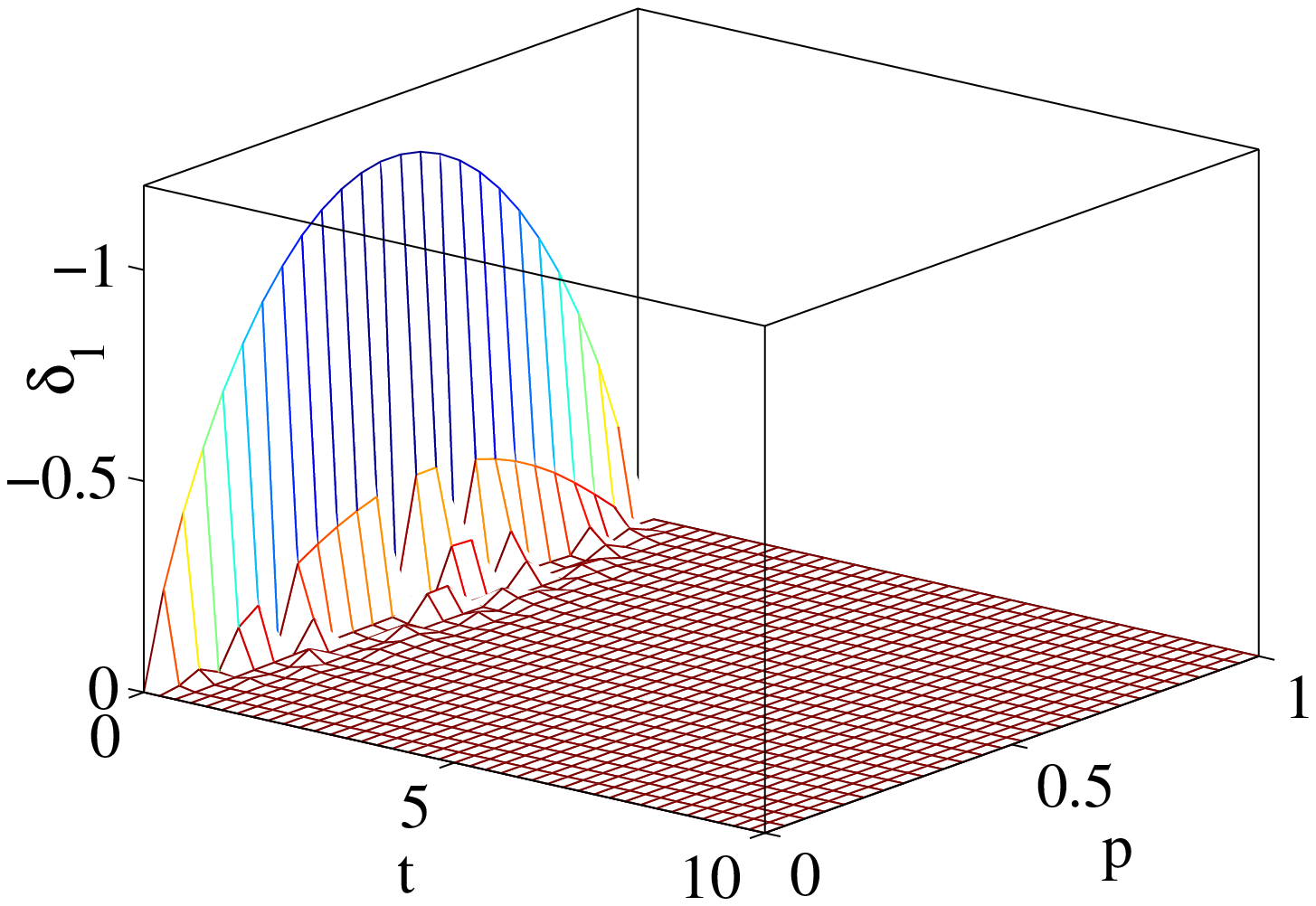}}
\subfigure{\includegraphics[width=4cm,height=4cm]{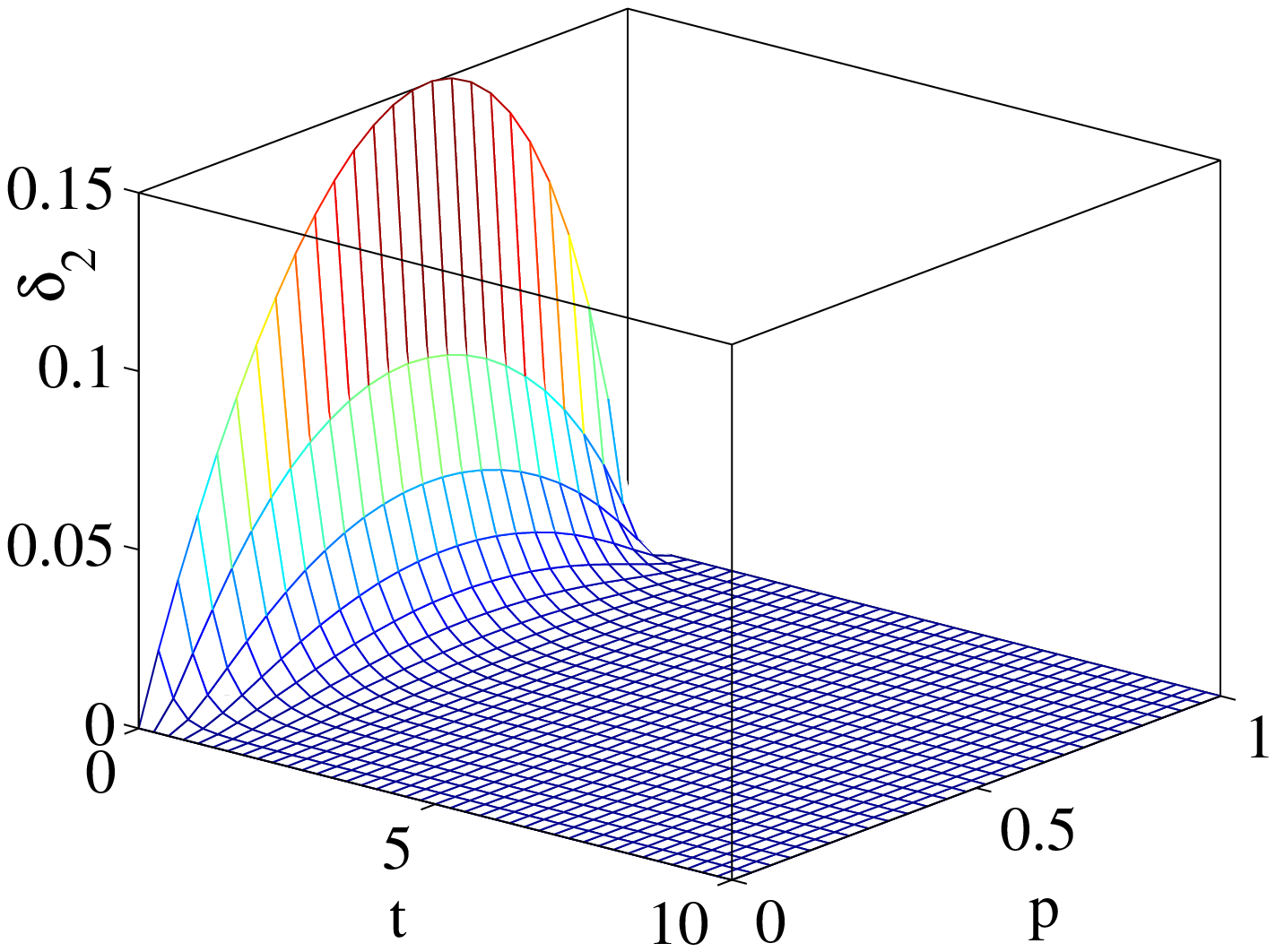}}
\caption{$\delta_{1}$ and $\delta_{2}$ dynamics of mixture of separable states $|000\rangle$ and $|+++\rangle$ in GAD Channel with $q=1/2$}
\end{center}
\end{figure}

\noindent \textit{\normalsize{4. Mixed Biseparable State-}}\\

\noindent We also studied the dynamics of the mixed biseparable state in presence of GAD 
channel for $q=1/2$ and we found that both dissensions remain at zero starting from the initial state.

\subsection{Effect of Dephasing Channel}
\noindent In this subsection, we consider the dephasing channel and its action on various three-qubit states. 
A dephasing channel causes loss of coherence without any energy exchange. The one-qubit Kraus operators for such 
process are given by $K_{0}$=diag (1,$\sqrt{1-\gamma}$) and $K_{1}$=diag(0,$\sqrt{\gamma}$).\\

\noindent \textit{\normalsize{1. Mixed GHZ State-}}\\

\noindent We once again consider the mixed GHZ state subjected to dephasing noise. The density matrix elements of the 
mixed GHZ at a time $t$ are given by,
\begin{eqnarray}
&&\rho_{11}=\rho_{88}=\frac{1}{8}(1+3p),\rho_{18}=\frac{p}{2}(1-\gamma)^\frac{3}{2},{}\nonumber\\&&
\rho_{ii}=\frac{1}{8}(1-p), i=2,...,7.\\
\end{eqnarray}
Here we observe that the diagonal elements are left intact whereas the off-diagonal elements undergo change as a 
consequence of dephasing noise. Interestingly, we find that $\delta_{1}$ is not at all influenced by dephasing 
channel whereas $\delta_{2}$ follows a regular asymptotic path [Fig.10]. The degradation observed in $\delta_{1}$ 
is due to progressively lower purity levels and is unaffected by dephasing noise. The reduced density 
matrices do not contribute towards monogamy score, thus making the dynamics of 
monogamy score just negative of $\delta_{2}$.\\
 
\begin{figure}[h]
\begin{center}
\subfigure{\includegraphics[width=4cm,height=4cm]{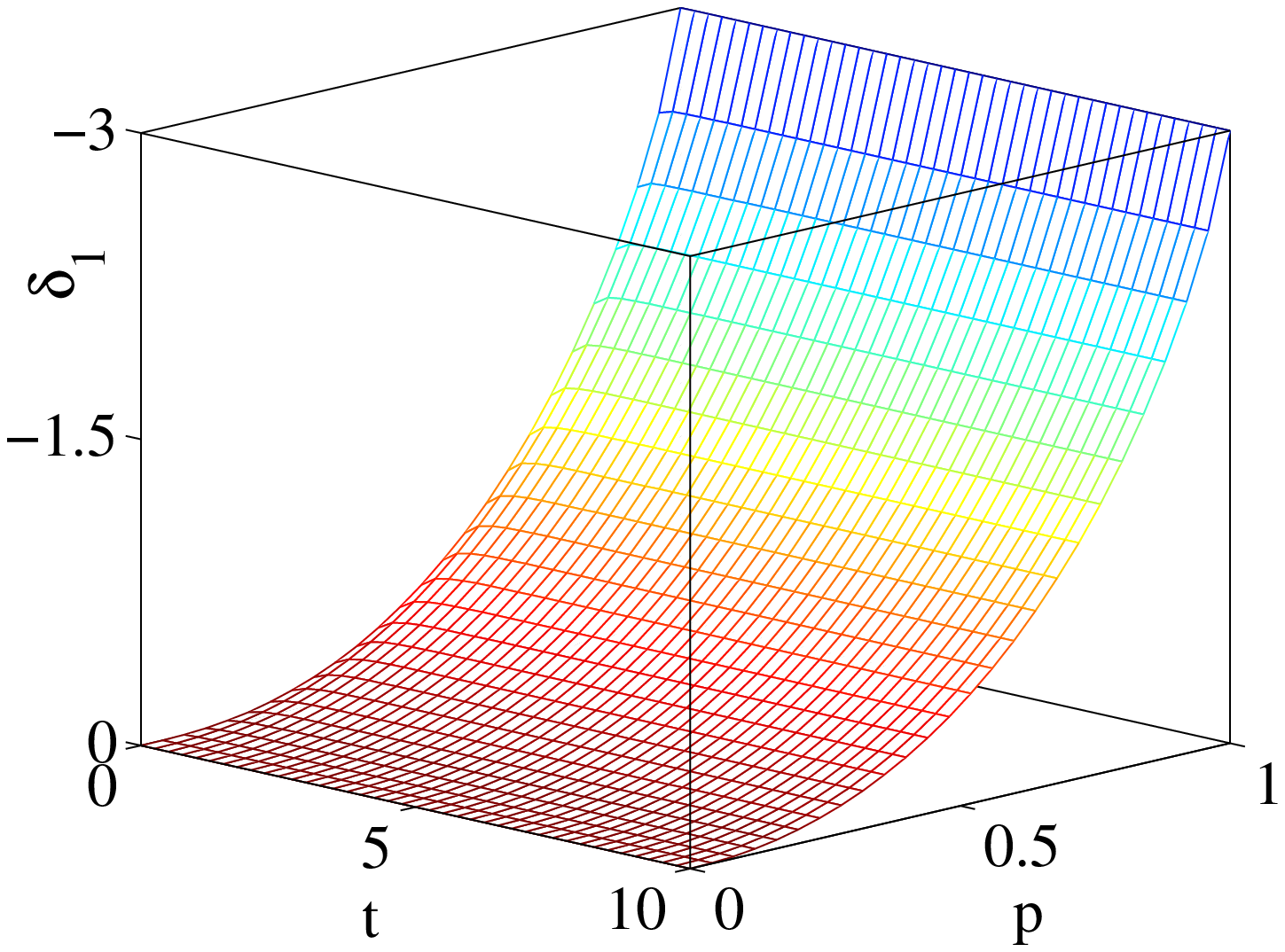}}
\subfigure{\includegraphics[width=4cm,height=4cm]{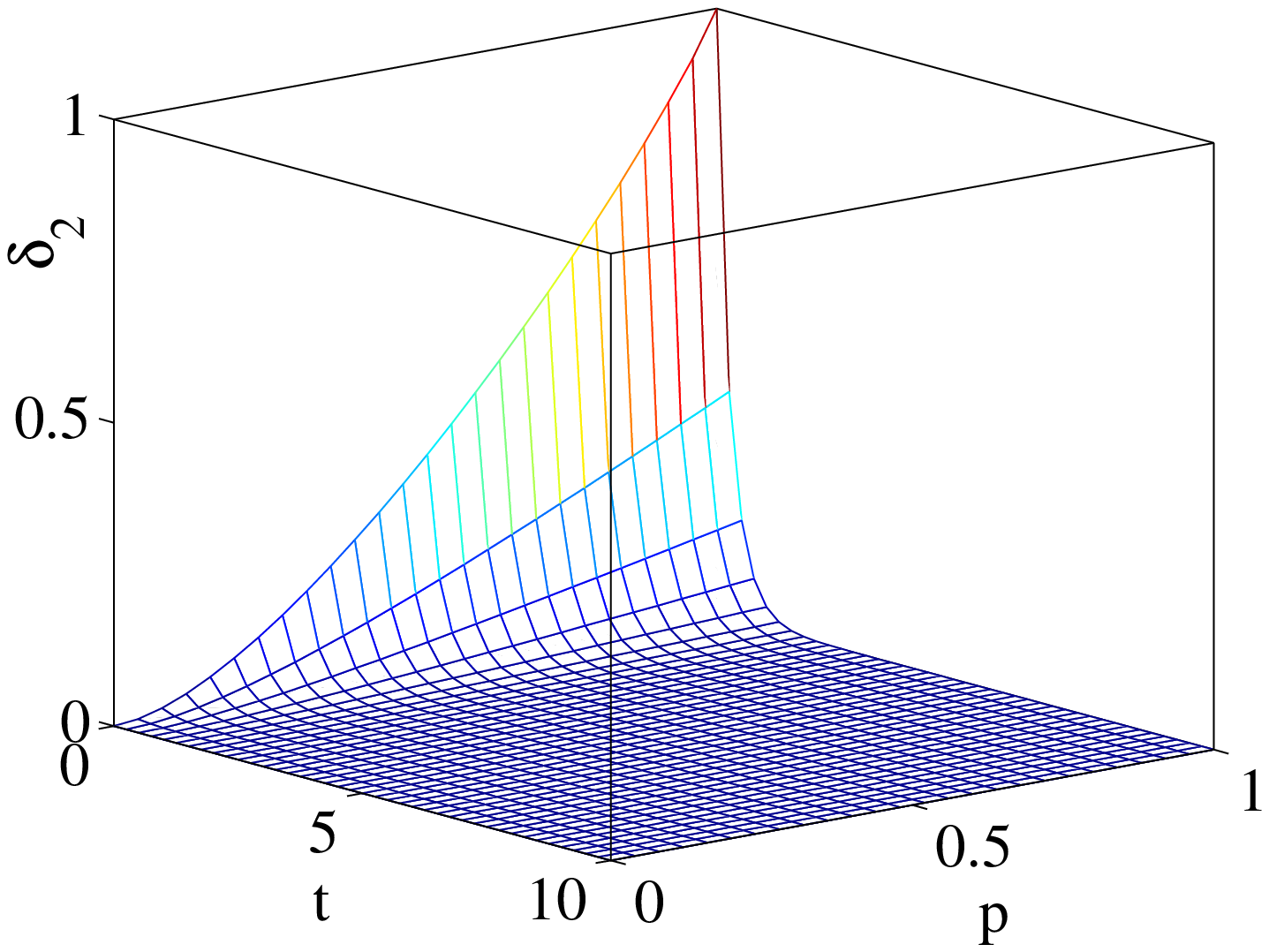}}
\caption{$\delta_{1}$ and $\delta_{2}$ dynamics of mixed GHZ state in dephasing channel.}
\end{center}
\end{figure}

\noindent \textit{\normalsize{2. Mixed W State-}}\\

\noindent The dynamics of mixed W state subjected to dephasing noise is as follows:
\begin{eqnarray}
&&\rho_{ii}=\frac{1}{8}(1-p) ,i=1,4,6,7,8,{}\nonumber\\&&
\rho_{22}=\rho_{33}=\rho_{55}=\frac{1}{24}(3+5p),{}\nonumber\\&&
\rho_{23}=\rho_{25}=\rho_{35}=\frac{1}{3}p(1-\gamma).
\end{eqnarray}
We noticed that for $p=1$, $\delta_{1}$ has a slower decay rate compared with other 
purity values and hence a finite amount of $\delta_{1}$ is present for all $t$ $\leq$ 10 at $p=1$ [Fig.11]. 
The decay of $\delta_{2}$ is asymptotic. For certain values of purity, the initial mixed W state is monogamous. 
However, they enter into the polygamous regime as a consequence of phase damping noise [Fig.12(a)]. 
After sufficient time, $\delta_{m}$ decays down to zero for all purity values.

\begin{figure}[h]
\begin{center}
\subfigure{\includegraphics[width=4cm,height=4cm]{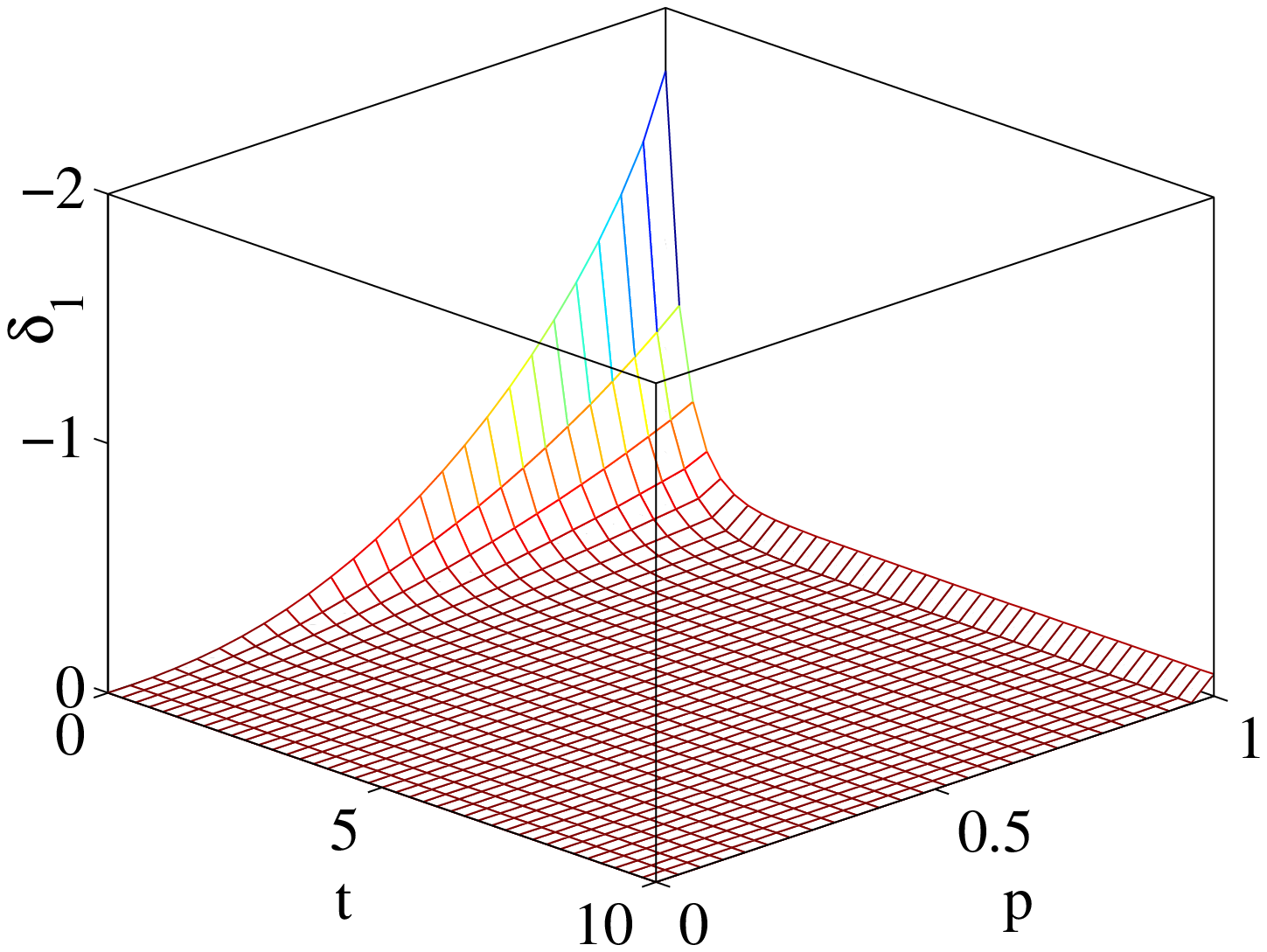}}
\subfigure{\includegraphics[width=4cm,height=4cm]{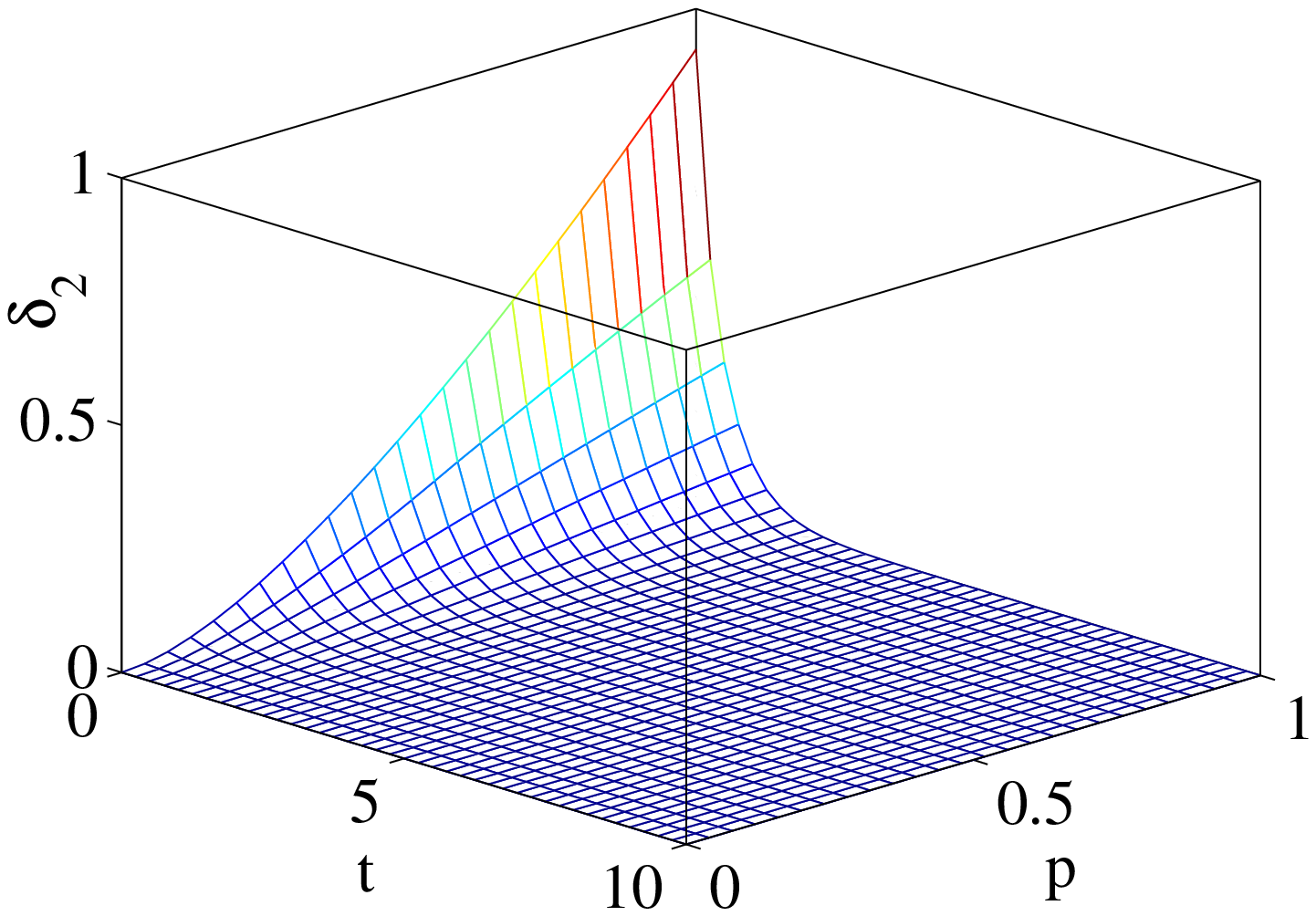}}
\caption{$\delta_{1}$ and $\delta_{2}$ dynamics of mixed W state in dephasing channel.}
\end{center}
\end{figure}

\begin{figure}[h]
\begin{center}
\subfigure{\includegraphics[width=4cm,height=4cm]{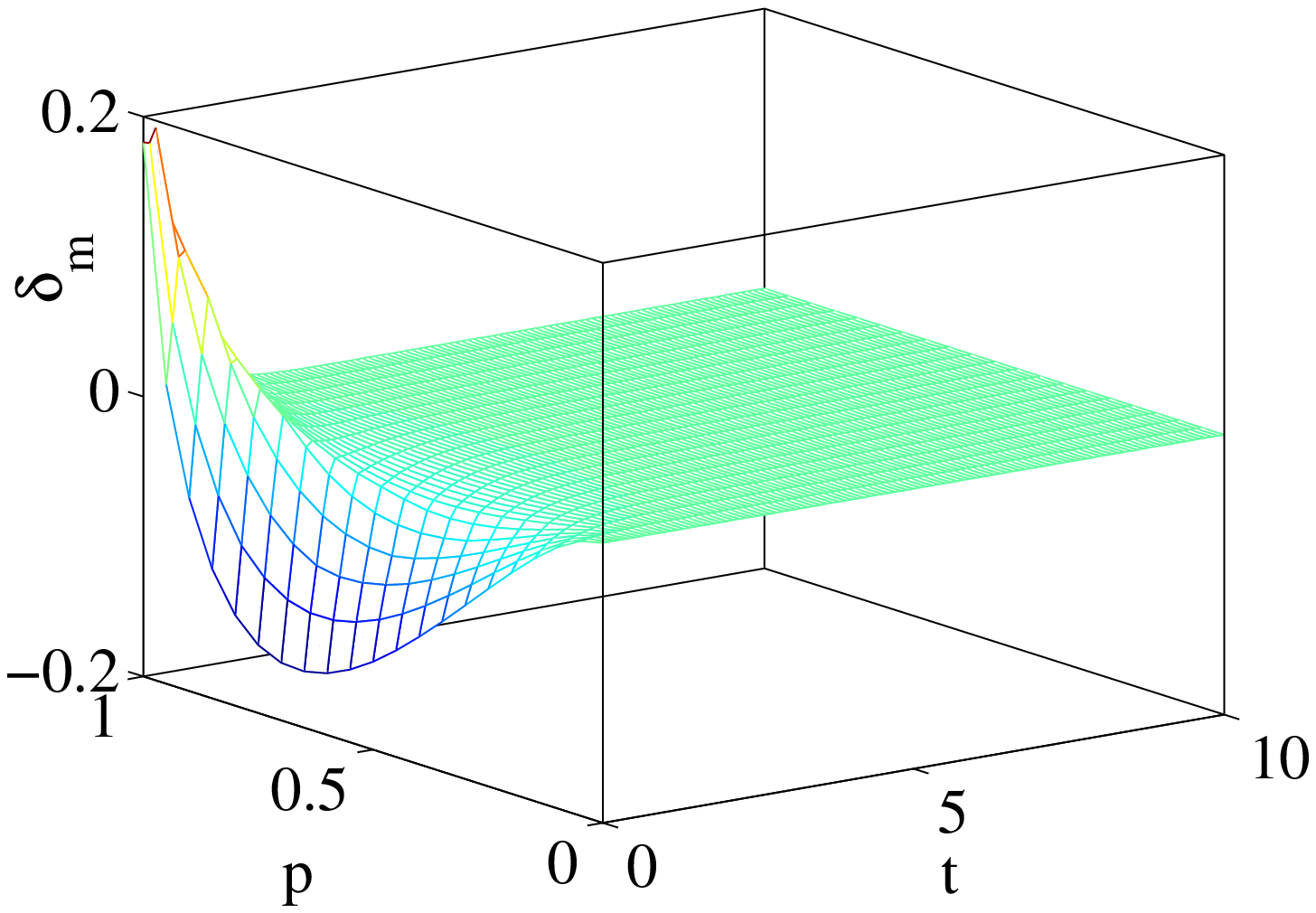}}
\subfigure{\includegraphics[width=4cm,height=4cm]{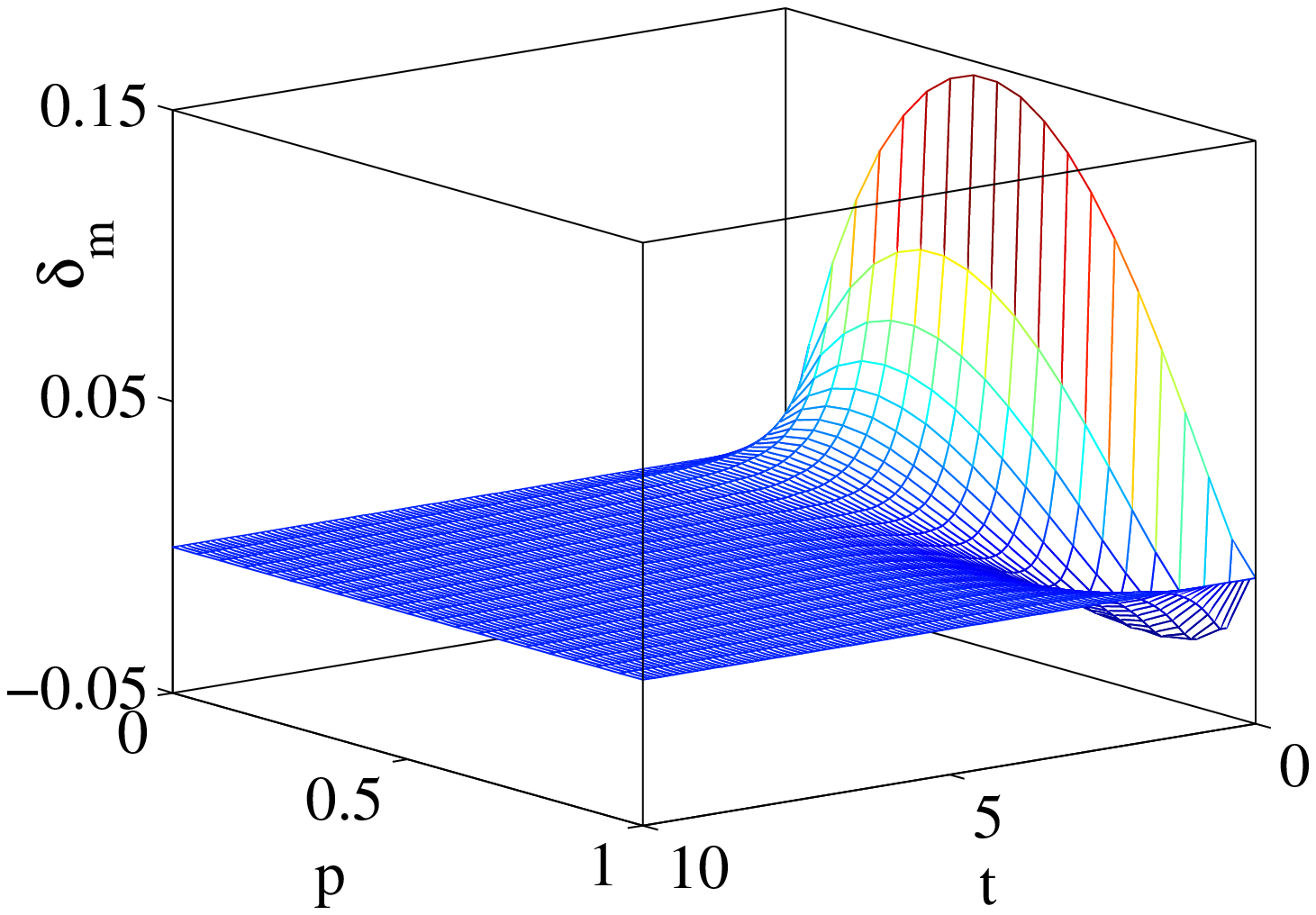}}
\caption{$\delta_{m}$ dynamics in dephasing channel of (a) Mixed W state, (b) Mixture of separable states: $p$$|$000$\rangle$$\langle$000$|$+$(1-p)$$|$+++$\rangle$$\langle$+++$|$}
\end{center}
\end{figure}

\noindent \textit{\normalsize{3. Mixture of Separable States-}}\\

\noindent The dynamics of $\rho=p|000\rangle\langle000|+(1-p)|+++\rangle\langle+++|$ under the influence of phase 
damping channel is given by:
\begin{eqnarray}
&&\rho_{11}=\frac{1}{8}(1+7p),\rho_{ii}=\frac{1}{8}(1-p) ,i=2,...,8,{}\nonumber\\&&
\rho_{12}=\rho_{13}=\rho_{15}=\rho_{24}=\rho_{26}=\rho_{34}=\rho_{37}=\rho_{48}={}\nonumber\\&&
\rho_{56}=\rho_{57}=\rho_{68}=\rho_{78}=\frac{1}{8}(1-p)\sqrt{1-\gamma},{}\nonumber\\&&
\rho_{14}=\rho_{16}=\rho_{17}=\rho_{23}=\rho_{25}=\rho_{28}=\rho_{35}=\rho_{38}={}\nonumber\\&&
\rho_{46}=\rho_{47}=\rho_{58}=\rho_{67}=\frac{1}{8}(1-p)(1-\gamma),{}\nonumber\\&&
\rho_{18}=\rho_{27}=\rho_{36}=\rho_{45}=\frac{1}{8}(1-p)(1-\gamma)^{\frac{3}{2}}.
\end{eqnarray}
Here, $\delta_{1}$ exhibits a strong revival all throughout the channel. 
However, the decay profile of $\delta_{2}$ is perfectly asymptotic [Fig.13]. Prior to channel action, 
i.e. at $t=0$, all density matrices are polygamous. With the action of the dephasing channel, density matrices 
with mixed ness closer to 1 enter into the monogamous regime [Fig.12(b)].\\

\begin{figure}[h]
\begin{center}
\subfigure{\includegraphics[width=4cm,height=4cm]{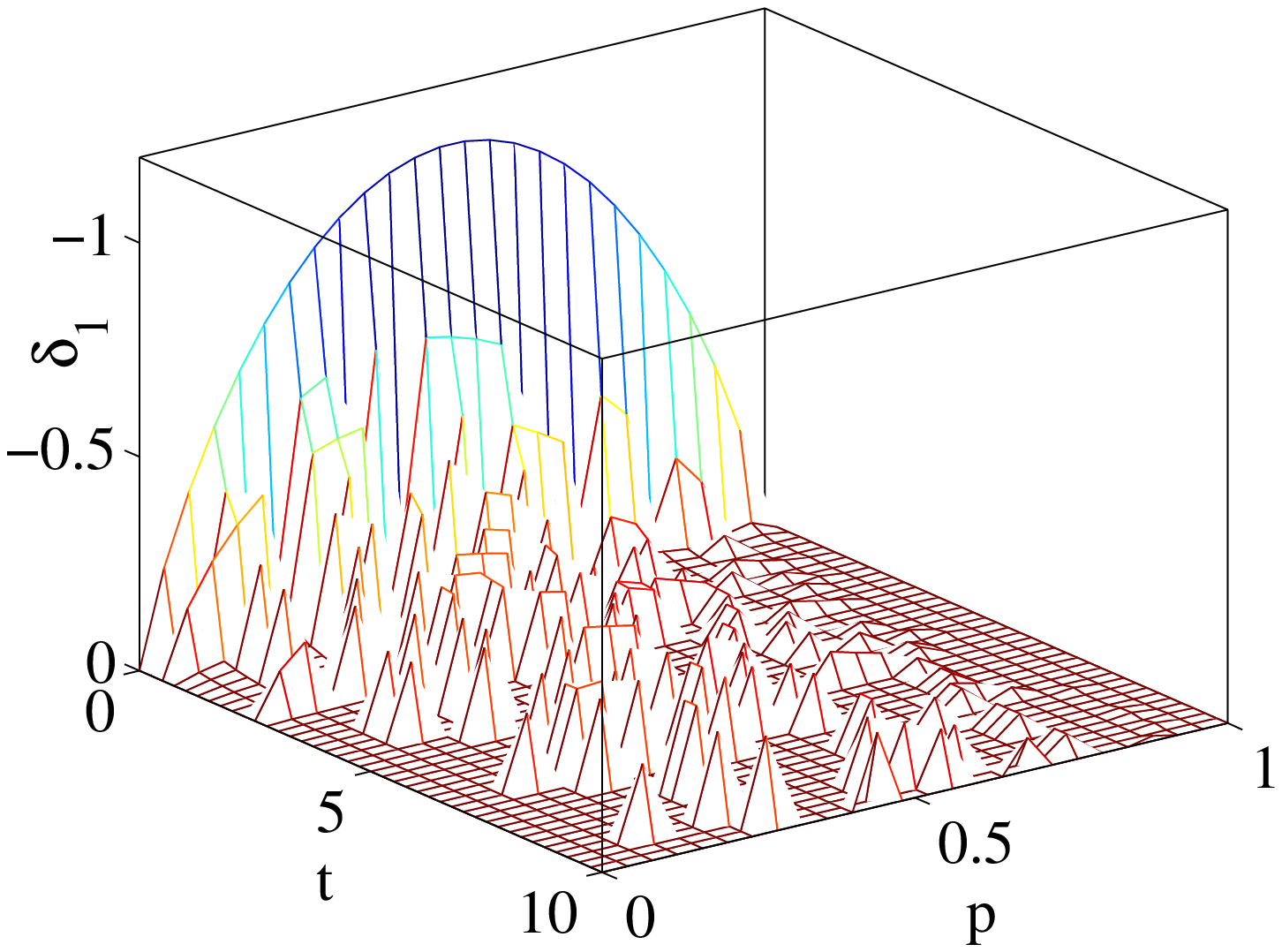}}
\subfigure{\includegraphics[width=4cm,height=4cm]{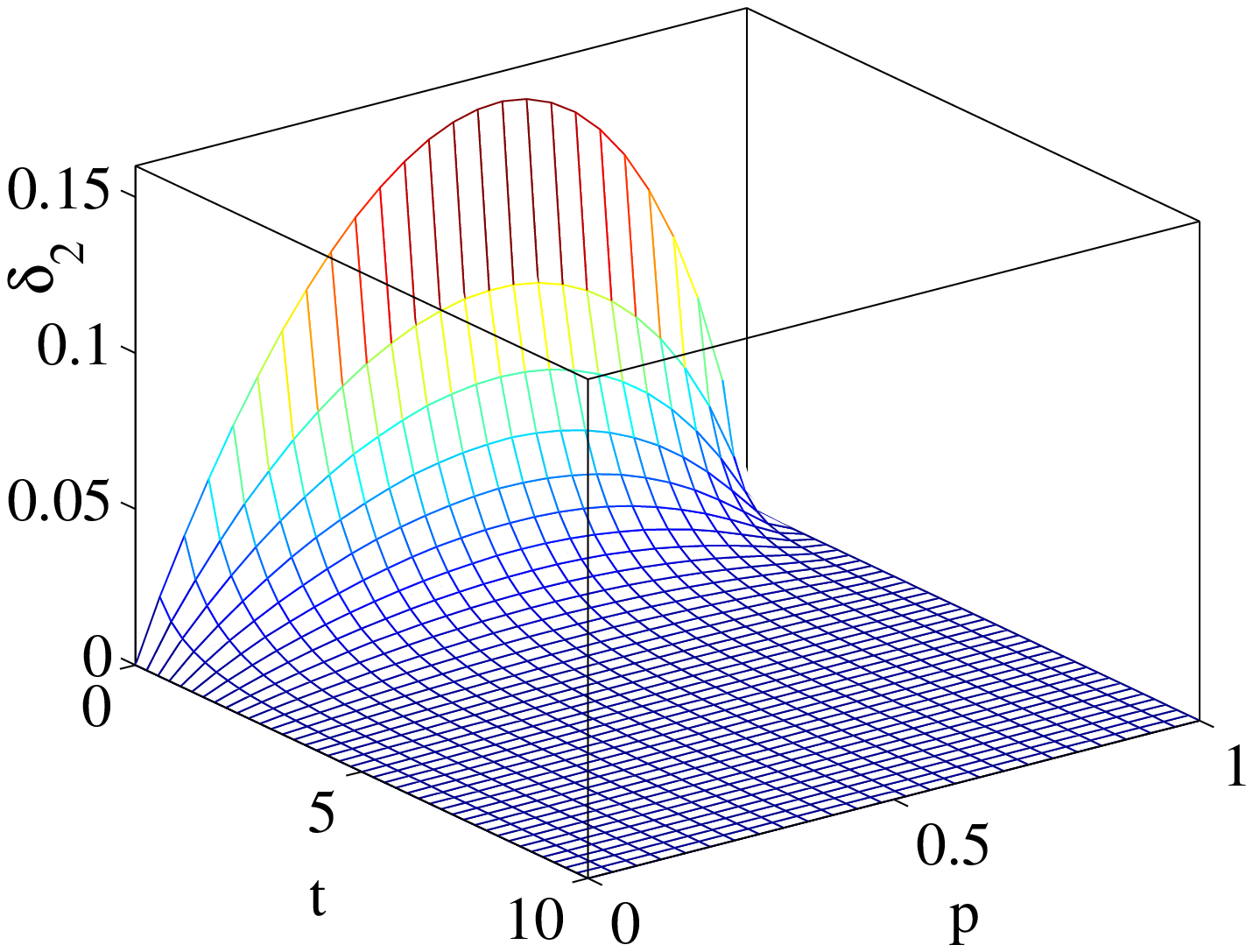}}
\caption{$\delta_{1}$ and $\delta_{2}$ dynamics of mixture of separable states $|000\rangle$ and $|+++\rangle$ in 
dephasing Channel}
\end{center}
\end{figure}

\noindent \textit{\normalsize{4. Biseparable State-}}\\

\noindent For the initial state, $\rho=(1-p)\frac{I}{8}+\frac{p}{2}[|0\rangle|\varphi^+\rangle\langle0|\langle\varphi^+|+|0\rangle|\psi^-\rangle\langle0|\langle\psi^-|$], the dynamics is given as:
\begin{eqnarray}
&&\rho_{ii}=\frac{1}{8}(1+p) ,i=1,2,3,4,{}\nonumber\\&&
\rho_{ii}=\frac{1}{8}(1-p) ,i=5,6,7,8,{}\nonumber\\&&
\rho_{14}=-\rho_{23}=\frac{p}{4}(1-\gamma).
\end{eqnarray}
Both $\delta_{1}$ and $\delta_{2}$ are zero throughout the channel operation time and do not 
show any revival.

\subsection{Effect of Depolarizing Channel}
\noindent In the final subsection of this section, we consider the effect of the depolarizing channel on three-qubit states. Under the action of a depolarizing channel, 
the initial single qubit density matrix dynamically evolves into a completely mixed state $I$/2. 
The Kraus operators representing depolarizing channel action are 
$K_{0}$=$\sqrt{1-3\gamma/4}$$\mathbb{I}$, $K_{1}$=$\sqrt{\gamma/4}\sigma_{x}$, $K_{2}$=$\sqrt{\gamma/4}\sigma_{y}$, 
$K_{3}$=$\sqrt{\gamma/4}\sigma_{z}$. (where $\sigma_{x}, \sigma_{y},\sigma_{z}$ are Pauli matrices)\\

\noindent \textit{\normalsize{1. Mixed GHZ State-}}\\

\noindent The dynamics of a mixed GHZ state when subjected to depolarizing channel is spelled out as,
\begin{eqnarray}
&&\rho_{11}=\rho_{88}=\frac{1}{8}[1+3p(1-\gamma)^2],\rho_{18}=\frac{p}{2}(1-\gamma)^3,{}\nonumber\\&&
\rho_{ii}=\frac{1}{8}[1-p(1-\gamma)^2]   i = 2,...,7.
\end{eqnarray}

\noindent Both $\delta_{1}$ and $\delta_{2}$ start decaying from the initial values of (-3.00,1.00) [Fig.14]. 
Quite interestingly, $\delta_{1}$ exhibits smooth asymptotic decay in contrary to the anomalies observed in 
case of $q=1/2$ GAD channel and dephasing channel. This instance underlines the fact that a certain noisy 
environment can largely influence the dynamics of multipartite quantum correlation. The depolarizing channel 
transfers the initial mixed GHZ state into $I$/8 which contains zero quantum dissension. Here the 
monogamy score $\delta_{m}$ of mixed GHZ state is just the negative of $\delta_{2}$.
\begin{figure}[h]
\begin{center}
\subfigure{\includegraphics[width=4cm,height=4cm]{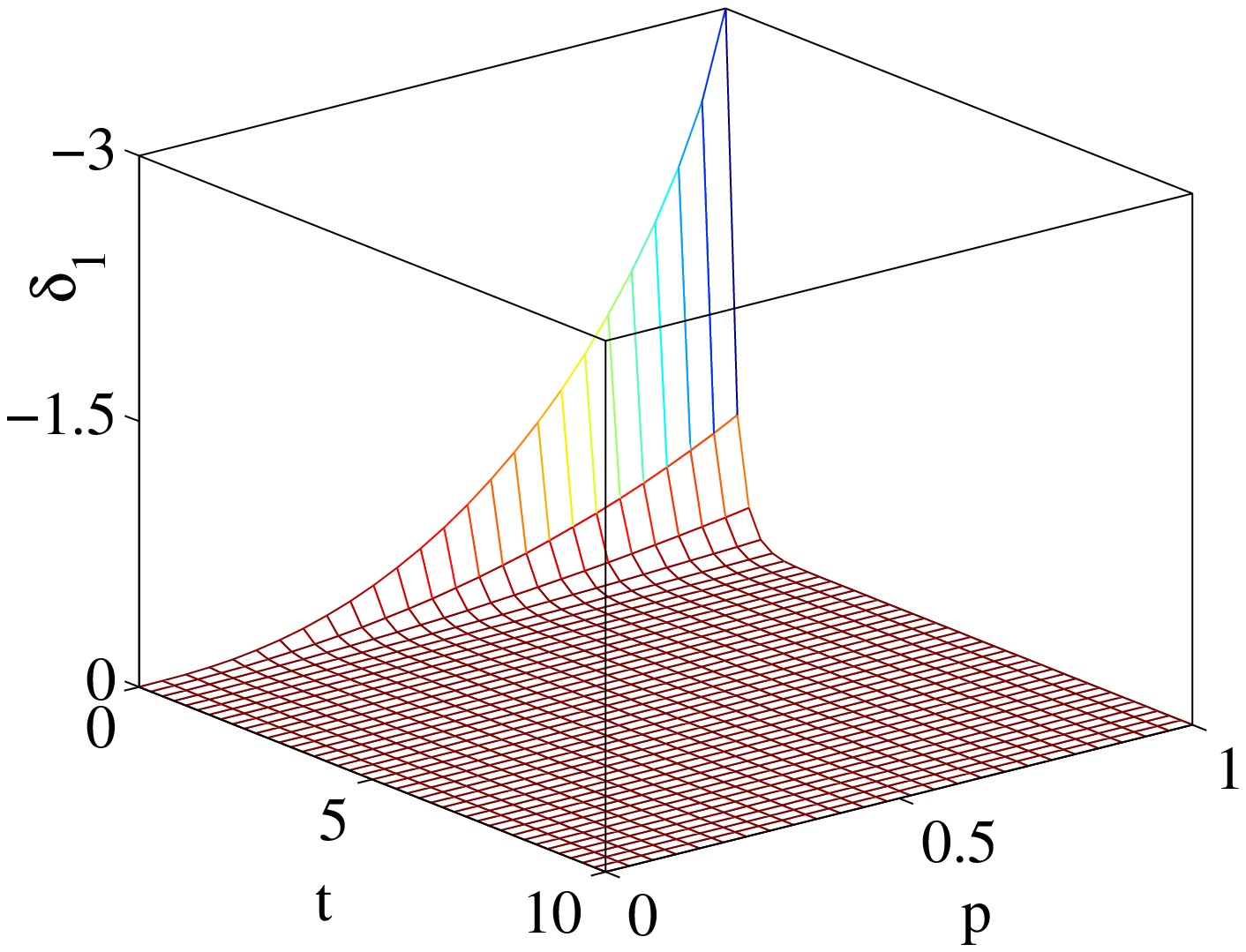}}
\subfigure{\includegraphics[width=4cm,height=4cm]{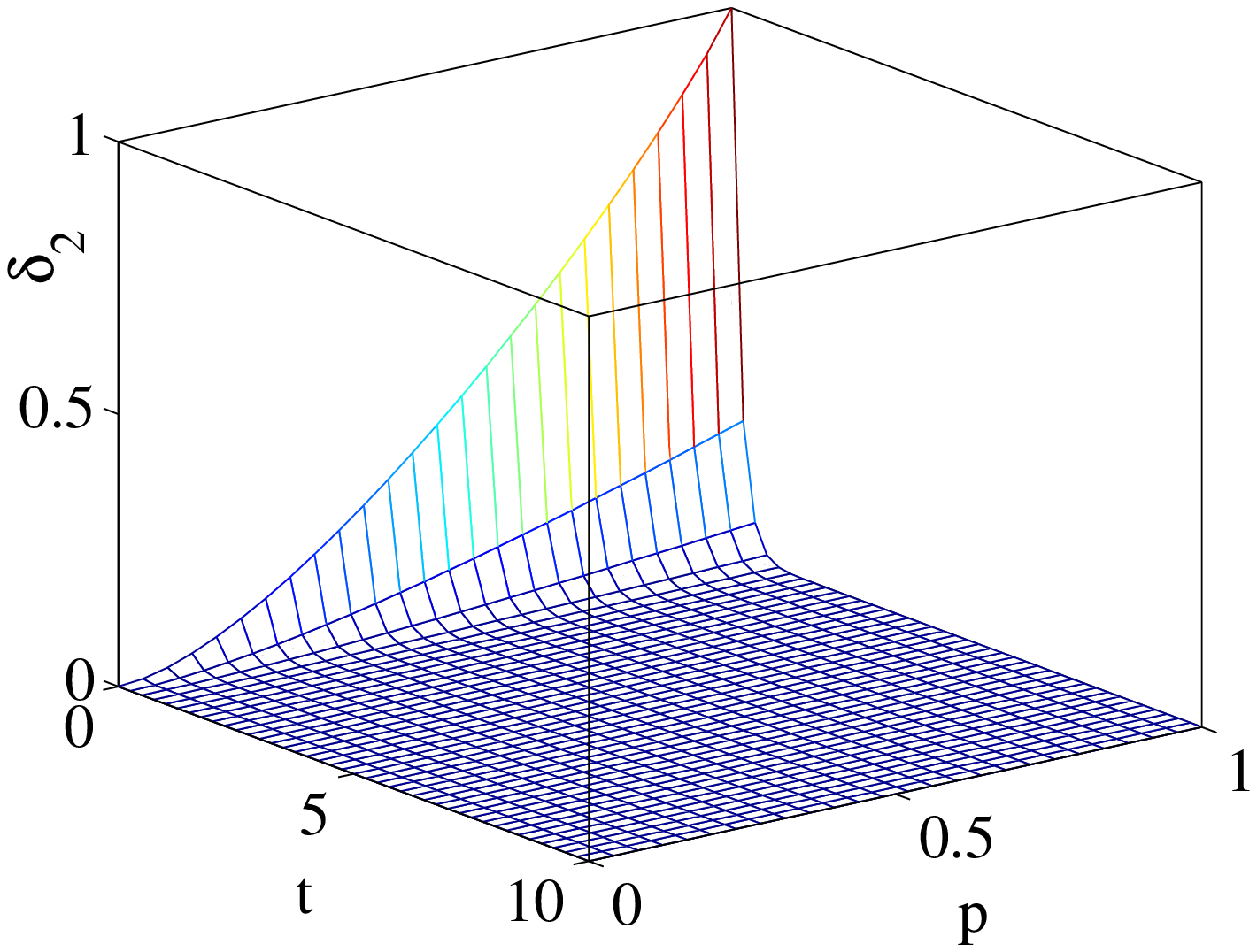}}
\caption{$\delta_{1}$ and $\delta_{2}$ dynamics of mixed GHZ state in depolarizing channel.}
\end{center}
\end{figure}

\noindent \textit{\normalsize{2. Mixed W State-}}\\

\noindent The dynamics of the mixed W state under the action of depolarizing channel is given by,
\begin{eqnarray}
&&\rho_{11}=\frac{1}{8}[1-p(1-\gamma)(\gamma^2-3\gamma+1)],{}\nonumber\\&&
\rho_{22}=\rho_{33}=\rho_{55}=\frac{1}{24}[3+p(1-\gamma)(3\gamma^2-7\gamma+5)],{}\nonumber\\&&
\rho_{23}=\rho_{25}=\rho_{35}=\frac{p}{6}(2-\gamma)(1-\gamma)^2,{}\nonumber\\&&
\rho_{44}=\rho_{66}=\rho_{77}=\frac{1}{24}[3-p(1-\gamma)(3\gamma^2-5\gamma+3)],{}\nonumber\\&&
\rho_{46}=\rho_{47}=\rho_{67}=\frac{p}{6}\gamma(1-\gamma)^2,{}\nonumber\\&&
\rho_{88}=\frac{1}{8}[1+p(1-\gamma)(\gamma^2-\gamma-1)].
\end{eqnarray}
\noindent Here, $\delta_{1}$ and $\delta_{2}$ attain maximum values of (-1.75,0.92) at $t=0$ and $p=1$ [Fig.15]. 
The initial mixed W state evolves to $I$/8 in the limit of $\gamma \rightarrow 1 $ resulting in zero quantum 
dissension. $\delta_{1}$ follows a perfect asymptotic path in contrast to the dynamics observed in case of 
$q=1/2$ GAD channel and dephasing channel. The monogamy score $\delta_{m}$ evolves as shown in Fig.16. 
For high purity values closer to 1, the initially polygamous states enter into monogamous regime owing to 
depolarizing channel action. On the other hand, states with low purity values which are initially monogamous 
do not experience any such transition.
\begin{figure}[h]
\begin{center}
\subfigure{\includegraphics[width=4cm,height=4cm]{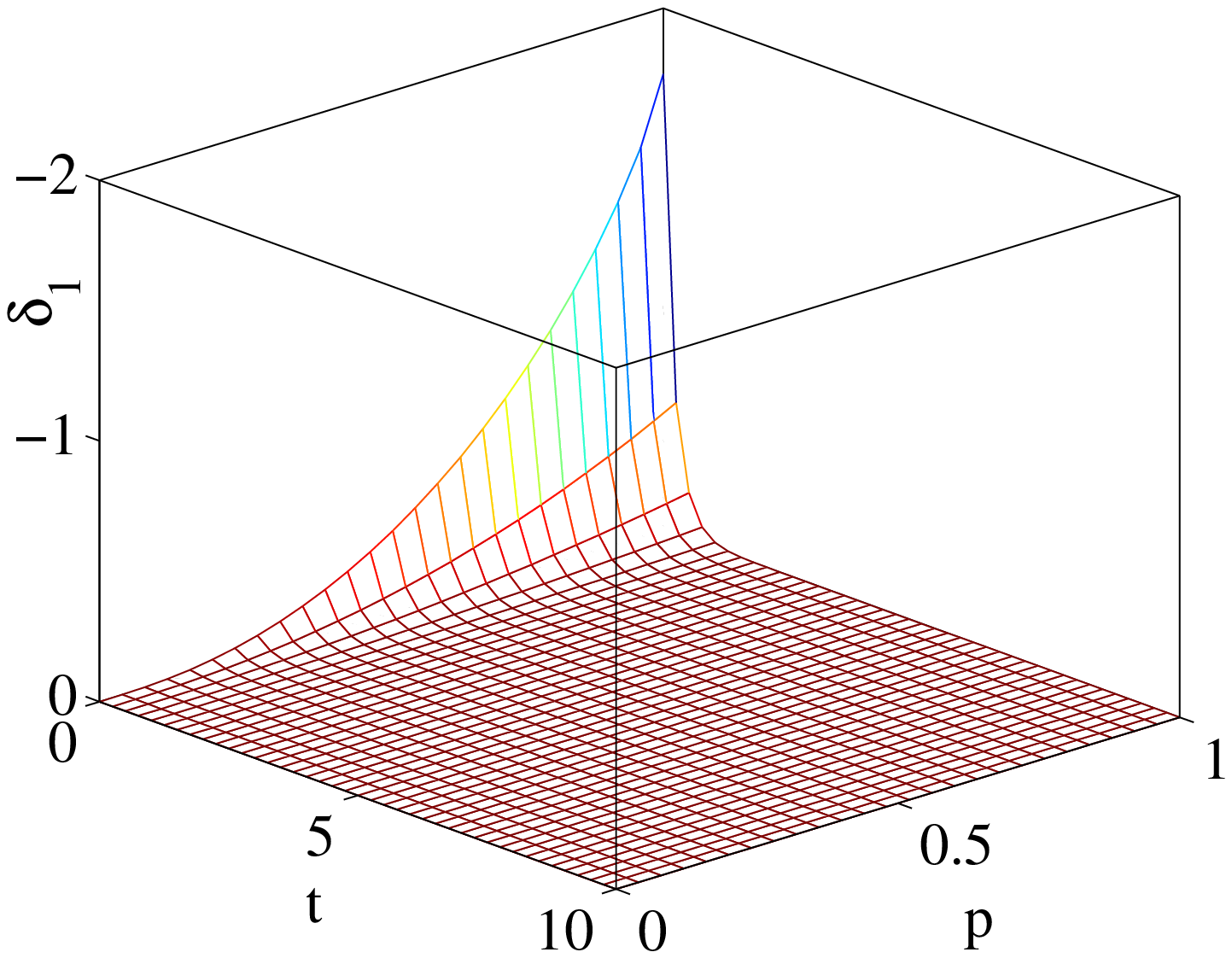}}
\subfigure{\includegraphics[width=4cm,height=4cm]{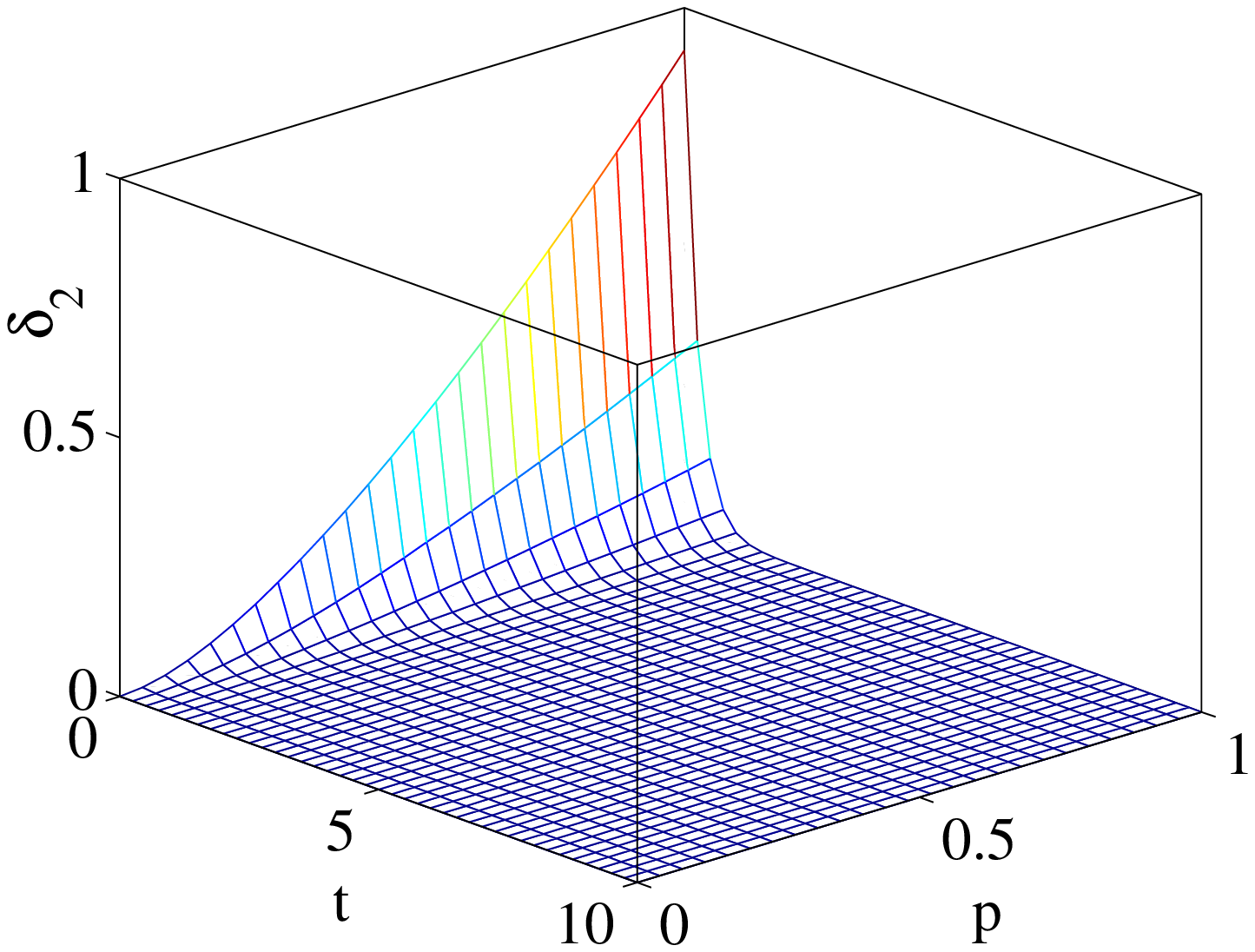}}
\caption{$\delta_{1}$ and $\delta_{2}$ dynamics of mixed W state in depolarizing channel.}
\end{center}
\end{figure}

\begin{figure}[h]
\begin{center}
\subfigure{\includegraphics[width=4cm,height=4cm]{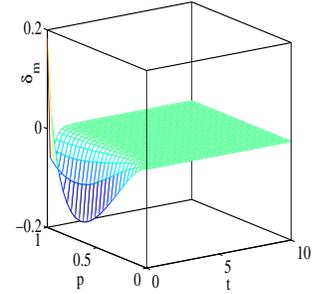}}
\caption{$\delta_{m}$ dynamics in depolarizing channel of  Mixed W state, (b) }
\end{center}
\end{figure}

\section{Conclusion}
\noindent In this work, we have extensively studied the dynamics of quantum correlation (quantum dissension) of 
various three qubit states like, mixed GHZ, mixed W, mixture of separable states and a mixed biseparable state 
 when these states are transferred through quantum noisy channels such as 
amplitude damping, dephasing and depolarizing. In most cases, we find that there is an asymptotic decay of 
quantum dissension with time. However, in certain cases, we have observed the revival of quantum correlation 
depending upon the nature of initial state as well as channel. This is quite interesting as we can explicitly see 
enhancement of multiqubit correlation in presence of local noise; similar in the line of quantum discord.\\
\noindent In addition, we have studied dynamics of monogamy score of three qubit states under different 
quantum noisy channels. Remarkably, we have seen that there are certain states which on undergoing effects of quantum 
channels change itself from monogamous to polygamous states. It is believed that monogamy property of the state is a 
strong signature of quantumness of the state and can be more useful security purpose compared to polygamous state. This 
study is useful from a futuristic perspective where we are required to create monogamous state from polygamous state 
for various cryptographic protocols.\\

\textit{Acknowledgment}
Authors acknowledge Prof A. K. Pati of Harish Chandra Research Institute for his invaluable suggestions in the 
improvement of the work. This work was initiated when the first author was at Indian Institute of Science.

\end{document}